\documentstyle[preprint,aps,epsf]{revtex}
\def\r#1{\ignorespaces $^{#1}$}
\def\mevcc {${\rm MeV}/c^2$}
\def\gevc {${\rm GeV}/c$}
\def\gevcc {${\rm GeV}/c^2$}
\def\ra   {\rightarrow}
\def\ckm  {Cabibbo-Kobayashi-Maskawa}
\newcommand{\Pt} {p_{\rm T}}
\newcommand{\Ds} {$D_{\mbox{\sf s}}^{-}$}
\newcommand{\Dsm} {D_{\mbox{\sf s}}^{-}}
\newcommand{\Bs} {$B_{\mbox{\sf s}}^{0}$}
\newcommand{\Bsm} {B_{\mbox{\sf s}}^{0}}
\newcommand{\Bsb} {\bar B_{\mbox{\sf s}}^{0}}
\newcommand{\Bsh} {B_{\mbox{\sf s}}^H}
\newcommand{\Bsl} {B_{\mbox{\sf s}}^L}
\newcommand{\Dsl} {D_{\mbox{\sf s}}^{-} \ell^+}
\newcommand{\lxy} {L_{\rm xy}}
\newcommand{\dgam} {\Delta\Gamma}
\newcommand{\dgog} {\Delta\Gamma/\Gamma}
\newcommand{\dm} {\Delta m}
\newcommand{\dms} {\Delta m_{\mbox{\sf s}}}
\newcommand{\phipi} {\phi \pi^-}
\newcommand{\kstark} {K^{*0} K^-}
\newcommand{\kstarpi} {K^{*0} \pi^-}
\newcommand{\ksk} {K^0_S K^-}
\newcommand{\phil} {\phi \mu^- \nu}
\newcommand{\Dsmu} {D_{\mbox{\sf s}}^{-} \mu^+} 
\newcommand{\dsd} {D_{\mbox{\sf s}} D}
\newcommand{\dsds} {D_{\mbox{\sf s}} D_{\mbox{\sf s}}} 
\newcommand{\fs} {f_{\mbox{\sf s}}}
\begin{document}
\tightenlines
\title{
\begin{flushright}
{\normalsize\rm 
Fermilab-Pub-98/172-E \\ 
\today}
\end{flushright}
Measurement of the \Bs\ Meson Lifetime \\ Using Semileptonic Decays
}
\author{
\font\eightit=cmti8
\hfilneg
\begin{sloppypar}
\noindent
F.~Abe,\r {17} H.~Akimoto,\r {39}
A.~Akopian,\r {31} M.~G.~Albrow,\r 7 A.~Amadon,\r 5 S.~R.~Amendolia,\r {27} 
D.~Amidei,\r {20} J.~Antos,\r {33} S.~Aota,\r {37}
G.~Apollinari,\r {31}
% ...
\end{sloppypar}
\vskip .026in
\begin{center}
(CDF Collaboration)
\end{center}
\newpage
\vskip .026in
\begin{center}
\r 1  {\eightit Argonne National Laboratory, Argonne, Illinois 60439} \\
\r 2  {\eightit Istituto Nazionale di Fisica Nucleare, University of Bologna,
I-40127 Bologna, Italy} \\
\r 3  {\eightit Brandeis University, Waltham, Massachusetts 02254} \\
\r 4  {\eightit University of California at Los Angeles, Los 
Angeles, California  90024} \\  
\r 5  {\eightit University of Chicago, Chicago, Illinois 60637} \\
\r 6  {\eightit Duke University, Durham, North Carolina  27708} \\
\r 7  {\eightit Fermi National Accelerator Laboratory, Batavia, Illinois 
60510} \\
\r 8  {\eightit University of Florida, Gainesville, Florida  32611} \\
\r 9  {\eightit Laboratori Nazionali di Frascati, Istituto Nazionale di Fisica
               Nucleare, I-00044 Frascati, Italy} \\
\r {10} {\eightit University of Geneva, CH-1211 Geneva 4, Switzerland} \\
\r {11} {\eightit Harvard University, Cambridge, Massachusetts 02138} \\
\r {12} {\eightit Hiroshima University, Higashi-Hiroshima 724, Japan} \\
\r {13} {\eightit University of Illinois, Urbana, Illinois 61801} \\
\r {14} {\eightit Institute of Particle Physics, McGill University, Montreal 
H3A 2T8, and University of Toronto,\\ Toronto M5S 1A7, Canada} \\
\r {15} {\eightit The Johns Hopkins University, Baltimore, Maryland 21218} \\
\r {16} {\eightit Institut f\"{u}r Experimentelle Kernphysik, 
Universit\"{a}t Karlsruhe, 76128 Karlsruhe, Germany} \\
\r {17} {\eightit National Laboratory for High Energy Physics (KEK), Tsukuba, 
Ibaraki 305, Japan} \\
\r {18} {\eightit Ernest Orlando Lawrence Berkeley National Laboratory, 
Berkeley, California 94720} \\
\r {19} {\eightit Massachusetts Institute of Technology, Cambridge,
Massachusetts  02139} \\   
\r {20} {\eightit University of Michigan, Ann Arbor, Michigan 48109} \\
\r {21} {\eightit Michigan State University, East Lansing, Michigan  48824} \\
\r {22} {\eightit University of New Mexico, Albuquerque, New Mexico 87131} \\
\r {23} {\eightit The Ohio State University, Columbus, Ohio  43210} \\
\r {24} {\eightit Osaka City University, Osaka 588, Japan} \\
\r {25} {\eightit Universita di Padova, Istituto Nazionale di Fisica 
          Nucleare, Sezione di Padova, I-35131 Padova, Italy} \\
\r {26} {\eightit University of Pennsylvania, Philadelphia, 
        Pennsylvania 19104} \\   
\r {27} {\eightit Istituto Nazionale di Fisica Nucleare, University and Scuola
               Normale Superiore of Pisa, I-56100 Pisa, Italy} \\
\r {28} {\eightit University of Pittsburgh, Pittsburgh, Pennsylvania 15260} \\
\r {29} {\eightit Purdue University, West Lafayette, Indiana 47907} \\
\r {30} {\eightit University of Rochester, Rochester, New York 14627} \\
\r {31} {\eightit Rockefeller University, New York, New York 10021} \\
\r {32} {\eightit Rutgers University, Piscataway, New Jersey 08855} \\
\r {33} {\eightit Academia Sinica, Taipei, Taiwan 11530, Republic of China} \\
\r {34} {\eightit Texas A\&M University, College Station, Texas 77843} \\
\r {35} {\eightit Texas Tech University, Lubbock, Texas 79409} \\
\r {36} {\eightit Istituto Nazionale di Fisica Nucleare, University of Trieste/
Udine, Italy} \\
\r {37} {\eightit University of Tsukuba, Tsukuba, Ibaraki 315, Japan} \\
\r {38} {\eightit Tufts University, Medford, Massachusetts 02155} \\
\r {39} {\eightit Waseda University, Tokyo 169, Japan} \\
\r {40} {\eightit University of Wisconsin, Madison, Wisconsin 53706} \\
\r {41} {\eightit Yale University, New Haven, Connecticut 06520} \\
\end{center}
} 	
\draft
\address{}
\date{\today}
\maketitle
\begin{abstract}
\noindent
The lifetime of the \Bs\ meson is measured using the semileptonic decay 
$\Bsm \ra \Dsl \nu X$.  The data sample 
consists of about $110$~pb$^{-1}$ of
$p \bar{p}$ collisions at $\sqrt{s} = 1.8$~TeV collected by the
CDF detector at Fermilab. Four different \Ds\ decay modes are
reconstructed resulting in approximately 600 $\Dsl$ signal events.
The \Bs\ meson lifetime is determined to be
$\tau(\Bsm) = (1.36 \pm 0.09 \ ^{+0.06}_{-0.05})$~ps,
where the first and second uncertainties are statistical and systematic,
respectively. 
The \Bs\ meson decay length distribution is examined
for a lifetime difference $\dgog$ between the two mass eigenstates of the
\Bs~meson.
An upper limit of $\dgog < 0.83$ is set at 95\% confidence level. 
\end{abstract}


\section{Introduction}

The lifetime differences between different bottom flavored hadrons
probe $B$~decay  
mechanisms which are beyond the simple quark spectator model. 
In the case of charm mesons, such differences have been observed to 
be quite large ($\tau(D^+)/\tau(D^0) \sim 2.5$)~\cite{PDG}.
Among bottom hadrons, the lifetime differences are expected to be 
smaller due to the heavier bottom quark mass \cite{bigi,neubert}. 
Some QCD inspired models based on the heavy quark expansion~\cite{bigi} 
predict a difference between  
the $B^+$ and $B^0$ meson lifetimes of about 5\% 
but expect the $B^0$ and \Bs\ lifetimes to differ by no more than 1\%.
Although some assumptions in Ref.~\cite{bigi} have been questioned 
in Ref.~\cite{neubert}, there is agreement that the
models expect a difference between the $B^0$ and \Bs\ lifetimes of
less than about 1\%.
These predictions are consistent with previous results of the
$B^0$ and $B^+$ meson lifetimes, as well as recent \Bs~lifetime 
measurements~\cite{PDG,bsall,old_bs_cdf}. 

In the Standard Model~\cite{GSW}, the \Bs~meson
exists in two $CP$-conjugate states, 
$|\Bsm\rangle = |\bar b s\rangle$ and
$|\Bsb\rangle = |b \bar s\rangle$.
The two mass eigenstates of the \Bs\ meson,
$\Bsh$ and $\Bsl$
($H =$ `heavy' and $L =$ `light'), are 
not $CP$ eigenstates but are mixtures of the two $CP$-conjugate quark
states:
\begin{equation}
|\Bsh\rangle = p\, |\Bsm\rangle - q\, |\Bsb\rangle \ {\rm and}\
|\Bsl\rangle = p\, |\Bsm\rangle + q\, |\Bsb\rangle, 
\ \ \ \ {\rm with}\ \ p^2 + q^2 = 1.
\end{equation}
The mass and lifetime differences between the $\Bsh$ and $\Bsl$ 
can be defined as
\begin{equation}
\Delta m \equiv m_H - m_L,\ \ \dgam \equiv \Gamma_L - \Gamma_H,\ \ 
{\rm and} \ \ \Gamma = \frac{\Gamma_H + \Gamma_L}{2},
\end{equation}
where $m_{H,L}$ and $\Gamma_{H,L}$ denote the mass and decay width of $\Bsh$
and $\Bsl$.
Unlike in the case of the $B^0$~meson, 
the width difference in the \Bs\ system is expected to be 
large~\cite{bigi_bs}.
Theoretical estimates 
predict $\dgog$ to be on the order of 10\% to 20\%~\cite{isi_bs,bbd_bs}.
In the \Bs\ system the ratio $\Delta m / \dgam$ is related to the ratio of the
\ckm~(CKM)~\cite{ckm} matrix 
elements $|V_{cb} V_{cs}| / |V_{ts} V_{tb} |$, which is quite well
known, and depends only   
on QCD corrections 
within the Standard Model~\cite{bbd_bs,datta}. 
Currently these QCD corrections are known to next-to-leading order in
the $1/m_b$ expansion~\cite{bbd_bs}.
A measurement of $\dgam$ would therefore imply a determination of $\dm$
and thus a 
way to infer the existence of \Bs\ meson oscillations, which will 
ultimately determine the ratio of the CKM matrix elements 
$|V_{td}| / |V_{ts}|$. 

It is assumed that \Bs\ mesons are produced 
as an equal mixture of $\Bsh$ and $\Bsl$~\cite{bbd_bs}.
In a search for $\dgam$, the \Bs~meson decay length 
distribution can be to described by a function of the form  
\begin {equation}
{\cal F}(t) = 
{\rm e}^{-\Gamma_H t} + {\rm e}^{-\Gamma_L t}\ \ \ \ \
{\rm with}\ \ \Gamma_{L,H} = \Gamma \pm \Delta\Gamma/2 ,
\end{equation}
rather than by just one exponential lifetime ${\rm e}^{-\Gamma t}$
which is the functional form used
in the measurement of the \Bs~lifetime assuming a single lifetime. 

In this paper, 
we present an update of the \Bs\ lifetime measurement at
CDF~\cite{old_bs_cdf} using the 
semileptonic decay\footnote{Throughout the paper
references to a specific  
charge state imply the charge-conjugate state as well.} 
$\Bsm \ra \Dsl \nu X$ ($\ell = e,\mu$), where the \Ds\ is identified via 
the four decay modes $\Dsm \ra \phipi$, $\kstark$, $\ksk$, and $\phil$.
We also examine the \Bs\ decay length distribution for a lifetime
difference $\dgog$ with a fit to two exponential lifetimes.
The data sample consists of approximately 110~pb$^{-1}$ of 
$p\bar{p}$~collisions at $\sqrt{s}$=1.8 TeV collected with the CDF
detector during Run~I. Of this, approximately 20~pb$^{-1}$ were collected
during the 1992-93 running period, while about 90~pb$^{-1}$ 
were accumulated during the 1994-96 run of the Tevatron Collider.
The result presented in this paper supersedes 
CDF's previous measurement of the \Bs\ lifetime using semileptonic
\Bs~decays~\cite{old_bs_cdf}.
That publication was based on 20~pb$^{-1}$ of data
and reconstructed the \Ds~meson in the $\phipi$ decay mode only.

The outline of this article is as follows: After a short description of
the CDF detector in Section~\ref{cdfdet}, the selection of the $\Dsl$
candidates is detailed in Section~\ref{datasel}.
The determination of the \Bs\ lifetime is the topic of
Section~\ref{lifeana}. We describe the search for a lifetime
difference $\dgog$ in Sec.~\ref{dgamogam} and offer
our conclusions in Section~\ref{conc}.

\section{The CDF Detector}
\label{cdfdet}

The Collider Detector at Fermilab (CDF) is a multi-purpose detector
designed to study 1.8~TeV $p \bar p$ collisions produced by the
Fermilab Tevatron Collider. 
The detector has a coordinate system with the $z$-axis along the
proton beam direction, the $y$-axis pointing vertically upwards, and
the $x$-axis pointing horizontally out of the Tevatron ring. 
Throughout this article $\varphi$ is the azimuthal angle, 
$\theta$ is the polar angle measured from the proton direction, 
and $r$ is the radius perpendicular to the beam axis. 
The CDF detector is described in detail elsewhere~\cite{cdf_det}. 
We summarize here only the detector features most relevant to this 
analysis. 

Three devices inside the 1.4 T solenoidal magnetic field 
are used for the tracking of charged particles: 
the silicon vertex detector (SVX), a set of vertex time projection
chambers (VTX), and the central tracking chamber (CTC).
The SVX \cite{cdf_svx} consists of four layers of silicon microstrip detectors 
located at radii between 2.9~cm and 7.9~cm from 
the interaction point. 
It provides spatial measurements in the $r$-$\varphi$ plane
with a track impact parameter resolution of about
$(13 + 40/\Pt)~\mu$m~\cite{cdf_svx}, 
where $\Pt$ is the component of the track
momentum~$p$ transverse to the $z$-axis ($\Pt = p \cdot \sin\theta$)
given in \gevc. 
The geometric acceptance of the SVX is about $60\%$ as it 
covers only $\pm~25$~cm from the nominal
interaction point whereas the luminous region of the Tevatron beam 
has an RMS of $\sim 30$~cm along the beam direction.

The VTX, which is located outside the SVX up to a radius of
22~cm, reconstructs track segments in the $r$-$z$ plane and is
used to determine the $z$-position of the primary interaction vertex
with a resolution of about 0.2~cm on average. Surrounding the SVX
and VTX is the CTC, located between radii of 30~cm and 132~cm. The
CTC is a 3.2~m long cylindrical drift chamber that contains 
84~layers of sense wires grouped into nine 
alternating super-layers of axial and stereo wires with a stereo angle
of $3^{\circ}$. 
The outer 54 layers of the CTC are instrumented to record the specific
ionization d$E$/d$x$ of charged particles. 
The CTC covers the pseudorapidity interval $|\eta|$ less than about 1.1, 
where $\eta=-\ln[\tan(\theta/2)]$.
The $\Pt$ resolution of the CTC combined with the SVX is 
$\sigma(\Pt)/\Pt = [(0.0066)^2 + (0.0009\,\Pt)^2]^{1/2}$, with $\Pt$
measured in \gevc.  

Outside the solenoid are electromagnetic (CEM) and hadronic (CHA)
calorimeters 
($|\eta|<1.1$) that employ a projective tower geometry
with a 
segmentation of $\Delta\eta \times \Delta\varphi \sim 0.1 \times 15^{\circ}$.
The sampling medium is composed of scintillators layered with lead and
steel absorbers.
A layer of proportional wire chambers (CES) 
is located near shower maximum in the CEM and 
provides a measurement of electromagnetic shower profiles 
in both the $\varphi$- and $z$-directions.

The muon detection system has four of its layers of planar drift
chambers (CMU) located beyond the central calorimeters.
To reduce the probability of misidentifying penetrating hadrons as muon
candidates in the pseudorapidity region $|\eta| \leq 0.6$,
four more layers of chambers (CMP) are located outside the magnet
return yoke. 
To reach these two detectors,
particles produced at the primary interaction vertex with a polar
angle of $90^{\circ}$ must traverse material totaling 5.4 and 8.4 pion
interaction lengths, respectively. 
An additional set of muon chambers (CMX) is located in the pseudorapidity
interval  $0.6 < |\eta| < 1.0$ to extend the polar acceptance of the
muon system.

\section{The Data Selection}
\label{datasel}

In this section, we describe the data selection, which begins with the
description of the lepton trigger data sets. This is followed by a
summary of the selection 
requirements, which are applied to
obtain the $\Dsl$ candidate events used for the \Bs\ lifetime
measurements. At the end of this section, we briefly describe the Monte
Carlo simulation of our data.

\subsection{The Lepton Trigger Data} 

Events containing semileptonic \Bs\ decays are collected using  
inclusive electron and muon trigger data as well as a data set obtained
from a dimuon trigger. CDF uses a three-level
trigger system. The first two levels are hardware based triggers, while
Level~3 is a software trigger based on the offline reconstruction code
optimized for computational speed.

At Level~1, inclusive electrons are selected by the presence of a single
calorimeter tower above a threshold of 6-8~GeV depending on run
conditions, while inclusive muons require 
the presence of a track in the CMU as well as the CMP. 
At Level~2, both of these triggers demand a charged track
with $\Pt>7.5$~\gevc\ reconstructed in the $r$-$\varphi$ plane of the CTC by
the central fast tracker (CFT)~\cite{cft}, 
a hardware track processor, which uses fast timing information from the
CTC as input. The momentum resolution of the CFT is
$\sigma(\Pt)/\Pt^2 = 3.5\%$ with a high efficiency.
In the case of the electron trigger, this track has to be matched to a
cluster in the electromagnetic calorimeter with transverse energy
$E_{\rm T}>8.0$~GeV, where $E_{\rm T}=E\cdot\sin\theta$, with $E$ being
the energy of the calorimeter cluster.
In the case of the muon trigger, this track must be
matched to a reconstructed track-segment in both the CMU and CMP.
At Level~3, a computer farm is used 
to fully reconstruct the data including three-dimensional track
reconstruction in the CTC. However, the fast algorithm used for
tracking is only efficient 
for particles with $\Pt>1.4$~\gevc.
In the third level of the trigger more stringent electron and muon
selection criteria, which are similar to those described in the next
Section~\ref{lepid}, are applied.

The reconstruction of the $\Dsm \ra \phil$ decay mode is based on
dimuon trigger data where the Level~1 trigger requires two muon
candidates be observed in the muon system. The second level trigger
requires the detection of at least one CFT track with $\Pt>2$~\gevc\
to match in $\varphi$ of each muon candidate. The third
level trigger requires that two reconstructed CTC tracks are matched
with two tracks in the muon chambers and the dimuon invariant mass is
less than 2.8~\gevcc. 
During Run~I about $7.5 \times 10^6$ electron
trigger events and about $2.5 \times 10^6$ inclusive muon trigger events
were recorded by CDF, while about 
$1.3 \times 10^6$ dimuon trigger events with
a dimuon invariant mass of less than 2.8 \gevcc\ were recorded.

\subsection{The Lepton Identification}
\label{lepid}

The identification of electron candidates reconstructed after data
collection uses information from both the
calorimeters and the tracking chambers. The
longitudinal shower profile has to be consistent with 
an electron
shower with a leakage energy from the CEM into the CHA of less than 4\% 
if one track is pointing to the calorimeter tower or less than 10\% if
more than one track is pointing to the calorimeter tower. The
lateral shower profile of the CEM cluster has to be 
consistent with that from test beam electrons. 
Additionally, a $\chi^2$ comparison of the CES shower profile with that
of test beam electrons has to result in $\chi^2 < 10$.
For the association of a single high $\Pt$
track with the calorimeter shower based on the position matching at
the CES plane, it is required that $r|\Delta \varphi| < 1.5$ cm and
$|\Delta z\sin \theta| < 3$~cm. In addition, we demand the $E_{\rm T}$
of the electron candidate reconstructed offline to be greater than 6~GeV. 
Electron candidates from photon conversion
due to detector material are reduced to less than 10\% by looking for
oppositely charged 
tracks which have a small opening angle with the electron
candidate. 

The reconstruction of muon candidates is described in
Ref.~\cite{lifetimeprd}. We compute a $\chi^2$ characterizing the
separation between the track segment in
the muon chamber and the extrapolated CTC track, where
the uncertainty in this $\chi^2$ variable is dominated by
the multiple scattering in the detector material. We require $\chi^2 < 9$
in the $r$-$\varphi$ view (CMU, CMP, and CMX) and $\chi^2 < 12$ in the
$r$-$z$ view (CMU and CMX). 
The transverse muon momentum
reconstructed offline is required to be $\Pt > 6$~\gevc. For the
dimuon sample this cut is $\Pt > 2$~\gevc\ for each muon candidate.
Finally, for optimal vertex 
resolution the electron and muon candidate tracks have to be
reconstructed in the SVX detector. 

\subsection{The \Ds\ Selection} 

The \Ds\ candidates are reconstructed in the decay modes 
\begin{itemize}
\item[(i)]   $\Dsm \ra \phipi$, $\phi \ra K^+ K^-$,
\item[(ii)]  $\Dsm \ra \kstark$, $K^{*0}  \ra K^+ \pi^-$,
\item[(iii)] $\Dsm \ra \ksk$, $K^0_S \ra \pi^+ \pi^-$,
\item[(iv)]  $\Dsm \ra \phil$, $\phi \ra K^+ K^-$.
\end{itemize}

The \Ds\ reconstruction is explained in the next section with the
example of the $\phipi$ 
decay channel. The other \Ds\ decay channels (ii)-(iv) are reconstructed
in a similar way. We then describe only the differences in the selection
of these decays (ii)-(iv) as compared to the $\Dsm \ra \phipi$ decay mode. 
The kinematic selection criteria used in this
analysis are optimized by maximizing the quantity $N_S / \sqrt{N_S + N_B}$, 
where $N_S$ is the predicted number of signal events based on Monte
Carlo calculations (see Sec.~\ref{mc}) and $N_B$ is the observed
number of background events estimated from the \Ds\ sideband regions
(see Sec.~\ref{fitter}).

\subsubsection{The $\Dsm \ra \phipi$ Mode}

The $\Dsm \ra \phipi$ reconstruction starts with a search 
for $\phi \ra K^+ K^-$ candidates. 
We first define a search cone around the lepton candidate with a 
radius $\Delta R = \sqrt{(\Delta\eta)^{2} + (\Delta\varphi)^{2}}$ of 0.8. 
Any two oppositely charged tracks 
with $\Pt >1.2$~\gevc\ within that cone are assigned the kaon mass and 
combined to form a $\phi$ candidate. 
Each $\phi$ candidate is required to have
a mass within $\pm 10$ MeV/$c^2$ of the world average 
$\phi$ mass~\cite{PDG}. The $\phi$ candidate is then 
combined with another track of $\Pt > 0.8$~\gevc\ inside the cone which 
has the opposite charge of the lepton (we call this the `right-sign'
combination). This  
third track is assigned the pion mass. To ensure a good decay vertex 
measurement, track quality cuts requiring a minimum number of hits in
the CTC are imposed on the 
tracks forming the \Ds\ candidate.
In addition, each of the tracks forming the \Ds\ is required to be
reconstructed in the SVX with hits in at least three out of the four silicon
layers and the $\chi^2$ of the track fit per SVX hit has to
be less than~6 to reject badly measured tracks. The same track
selection requirements are also applied to the lepton candidate tracks.

The specific ionization information d$E$/d$x$ from the CTC is used to
help identify hadrons in the \Ds\ reconstruction. Because of the large
Landau tail of the ionization distribution, the 80\% truncated mean
of the measured charges from the CTC sense wires is
taken as the best estimator of the track d$E$/d$x$. The probabilities,
$P(i)$, for
a track to be consistent with the $i = e$, $\mu$, $\pi$, $K$, or $p$
hypotheses are then calculated using the measured d$E$/d$x$ value and
the predictions for the assumed particle hypotheses. We define a
likelihood ratio, 
$\ell h_{{\rm d}E/{\rm d}x}^K$, 
for a track being a
kaon to be the ratio of $P(K)$ divided by the
sum of the probabilities of all particle hypotheses. 
The quantity $\ell h_{{\rm d}E/{\rm d}x}^{\pi}$ 
is defined correspondingly. 
We require the likelihood ratios $\ell h_{{\rm d}E/{\rm d}x}^K > 0.01$ and  
$\ell h_{{\rm d}E/{\rm d}x}^{\pi} > 0.01$. 
  
Since the $\phi$ has spin 1 and both the \Ds\ and $\pi^-$ are spin 0, the 
helicity angle $\Psi$, which is the angle between the $K^-$ and \Ds\ 
directions in the $\phi$ rest frame, exhibits a distribution 
$dN/d(\cos\Psi) \sim \cos^2\Psi$. A cut $|\cos \Psi| > 0.4$ is 
therefore applied to suppress combinatorial background, which 
is found to be flat in the $\cos\Psi$ distribution. 
We also apply an 
isolation cut $E_{\rm T}^{\rm iso}/\Pt (\phi\pi^-) < 1.0$ on the \Ds\ 
candidate, where $E_{\rm T}^{\rm iso}$ is the sum of the transverse
energy within a cone of  
radius 0.4 in $\eta$-$\varphi$ space around the lepton candidate,
excluding the lepton energy.   
This cut eliminates many of the fake \Ds\ combinations from high 
track multiplicity jets.  
Finally, the mass of the
$\Dsl$ system is required to be between 3.0~\gevcc\ and 5.0~\gevcc\ 
to be consistent with coming from a \Bs\ decay. 

\begin{figure}[tbp]\centering
\begin{picture}(160,180)(5,-5)
\put(19,74){\large\bf (c)}
\put(106,74){\large\bf (d)}
\put(19,161){\large\bf (a)}
\put(106,161){\large\bf (b)}
\centerline{
\epsfysize=17.0cm
\epsffile[60 185 540 655]{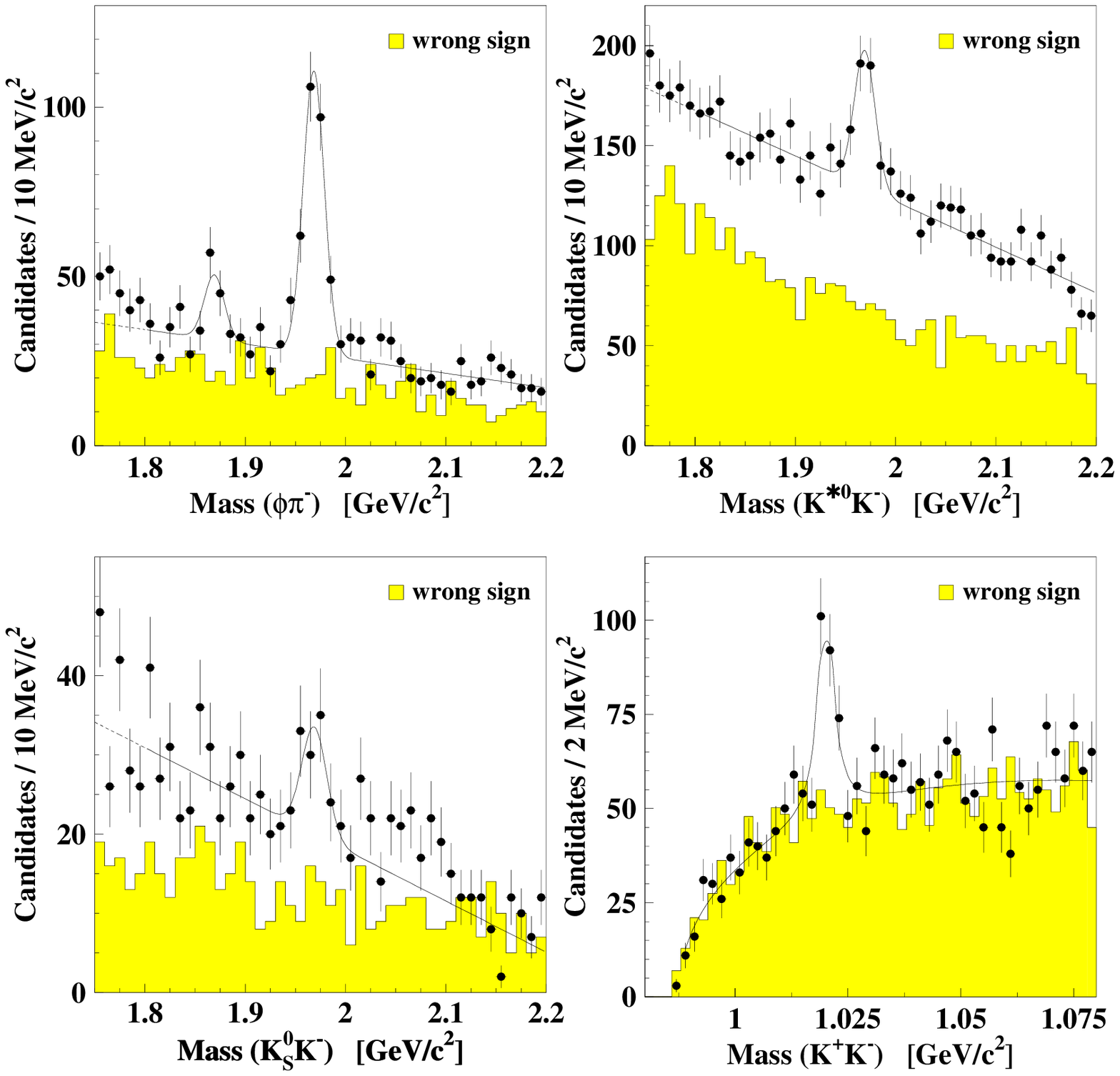}
}
\end{picture}
\caption{Invariant mass distributions of 
(a) $\Dsm \ra \phipi$, (b) $\Dsm \ra \kstark$,
(c) $\Dsm \ra \ksk$, and (d) $\phi \ra K^+K^-$ from $\Dsm \ra \phil$. 
The dots with error bars are 
for `right-sign' $\Dsl$ combinations while the
shaded histograms show
the corresponding `wrong-sign' distributions.
In (a) evidence of the decay $D^- \ra \phi \pi^-$ is present.
The results of the fits described in the text are also
superimposed. The mass regions indicated by a dashed line have not
been included in the fits.}
\label{ds_mass}
\end{figure}

In Figure~\ref{ds_mass}a) the $\phipi$ invariant mass distribution 
for the `right-sign' $\Dsl$ combinations is displayed. A \Ds\ signal 
of $220 \pm 21$ events 
is observed from fitting a Gaussian with a mean of 
$(1968 \pm 1)$~\mevcc\ and a width of $(10.5 \pm 1.1)$~\mevcc\ plus a
straight line to describe the combinatorial background. 
The mass region indicated by a dashed line has not
been included in the fit to avoid contributions from \Ds\ decays 
where particles have been missed such as the $\pi^0$ in the decay
$\Dsm \ra \phipi \pi^0$.  Evidence
of the Cabibbo-suppressed decay $D^- \ra \phi \pi^-$ is also 
present in Fig.~\ref{ds_mass}a). The shaded distribution shows
$\phipi$ candidates from 
`wrong-sign' $\Dsm\ell^-$ combinations.
Here, no enhancement is seen in the \Ds\ mass region. 
We allow multiple $\phipi$~combinations per event but the number of
multiple entries is found to be less than 1\% of the total number of
combinations. This is also the case for 
the $\Dsm \ra \kstark$ and $\ksk$ decay modes.

\subsubsection{The $\Dsm \ra \kstark$ Mode}

For the $\Dsm \ra \kstark$ decay mode with $K^{*0} \ra K^+ \pi^-$, 
we assign the $K^+$ and
$\pi^-$ masses to two oppositely charged tracks found in the cone of
$\Delta R < 0.8$ around the lepton. To reflect the different decay
kinematics of the 
$K^{*0} \ra K^+ \pi^-$ decay compared to $\phi \ra K^+K^-$, we
require $\Pt(K^+) > 1.2$~\gevc\ and $\Pt(\pi^-) > 0.4$~\gevc. 
Each $K^{*0}$ candidate is required to have
a mass within $\pm 40$ MeV/$c^2$ of the world average 
$K^{*0}$ mass~\cite{PDG}. To further reduce the high combinatorial
background in this decay channel, we tighten the helicity cut to
$|\cos \Psi| > 0.5$ and introduce a track based isolation requirement
$\Pt(\Dsl)/\Pt({\rm cone}) > 0.6$, where $\Pt({\rm cone})$ is the sum of the
transverse momenta of all tracks in a cone $\Delta R < 1.0$ in
$\eta$-$\phi$ space. The cone has the lepton direction as its axis and
the primary event vertex (see Sec.~\ref{primvx}) as its vertex.
All other selection requirements discussed in the
$\phipi$ decay mode remain the same with the exception of the 
$\ell h_{{\rm d}E/{\rm d}x}^{K^-}$ cut described below.

The $\kstark$ invariant mass distribution for the `right-sign' $\Dsl$
combinations is shown as dots with error bars in Figure~\ref{ds_mass}b).
A signal at the \Ds\ mass is visible. This signal 
contains also events from a $D^- \ra K^{*0} \pi^-$ reflection, where the
$\pi^-$ is incorrectly assigned the kaon mass. This reflection is further
discussed in Section~\ref{reflect}. To reduce the effect of this
reflection the d$E$/d$x$ requirement for that track 
is tightened to
$\ell h_{{\rm d}E/{\rm d}x}^{K^-} > 0.1$, 
while we still demand 
$\ell h_{{\rm d}E/{\rm d}x}^{\pi,K} > 0.01$ for the tracks forming the
$K^{*0}$ candidate.
 
\subsubsection{The $\Dsm \ra \ksk$ Mode}

The reconstruction of the $\Dsm \ra \ksk$ decay mode begins with a
search for $K^0_S \ra \pi^+\pi^-$ candidates by assigning the pion
mass to any two oppositely charged tracks with $\Pt > 0.4$~\gevc\ in a
cone of $\Delta R < 1.0$ in $\eta$-$\phi$ space around the lepton. These
two tracks are constrained to come from a common vertex and their
invariant mass has to be within $5\sigma$ of the nominal $K^0_S$
mass~\cite{PDG}, where $\sigma$ is the uncertainty on the 
$\pi^+\pi^-$ mass measurement. 
Exploiting the long lifetime of the $K_S^0$ meson, we require the
$K_S^0$ vertex to be significantly displaced from the primary event
vertex, which is further described in Sec.~\ref{primvx}. 
We determine the transverse decay
length $\lxy$ (see Sec.~\ref{decaylength}) of the $K_S^0$ and require
$\lxy > 3 \sigma$, where $\sigma$ is the measured uncertainty on $\lxy$
for each candidate event.

The $K^0_S$ candidate is combined with
any kaon candidate with $\Pt > 1.2$~\gevc\ within $\Delta R < 0.8$
around the lepton to find the \Ds\ candidate. 
The dots with error bars in Figure~\ref{ds_mass}c) show the
$\ksk$ invariant mass distribution for the `right-sign' $\Dsl$
combinations. An enhancement at the \Ds\ mass is visible. 
As in the $\kstark$ mode, this signal 
contains events from a $D^- \ra K^0_S \pi^-$ reflection, where the
$\pi^-$ is assigned the kaon mass (see 
Section~\ref{reflect}). To reduce the effect of this
reflection, we again require
$\ell h_{{\rm d}E/{\rm d}x}^{K^-} > 0.1$, 
while we demand
$\ell h_{{\rm d}E/{\rm d}x}^{\pi} > 0.01$ for the tracks forming the
$K^0_S$.

\subsubsection{The $\Dsm \ra \phil$ Mode}

For the $\Dsm \ra \phil$ decay mode, we start with two
oppositely charged muons with $\Pt > 2$~\gevc\ utilizing the dimuon
data set obtained with a trigger which requires the dimuon 
invariant mass to be smaller than 2.8~\gevcc. This requirement is more than
90\% efficient for a double semileptonic decay $\Bsm \ra \Dsmu \nu X$
followed by $\Dsm \ra \phil$. In addition,
two oppositely charged tracks 
with $\Pt > 0.8$~\gevc\ are assigned the kaon mass and 
combined to form a $\phi$ candidate. 
There is an ambiguity in the assignment of one of the two muons to a
found $\phi\ra 
K^+K^-$ candidate. One of the muons comes from the \Ds\ semileptonic
decay ($\mu_{\Dsm}$) while the other originates from the \Bs\ decay
($\mu_{\Bsm}$). 
In order to resolve this ambiguity, we require m$(KK\mu_{\Dsm})$ to be
smaller than the world average \Ds~mass~\cite{PDG}, while
m$(KK\mu_{\Bsm})$ has to be greater than m$_{\Dsm}$.
To reduce combinatorial
background in this decay channel, we use
the track based isolation quantity 
$\Pt(\Dsl)/\Pt({\rm cone})$ and require it to be greater than 0.5.
As required in the other decay modes, the invariant mass of the
$K K \mu \mu$ system has to lie between 3.0~\gevcc\ and 5.0~\gevcc.
The number of multiple $\Dsm\mu^+$ combinations per event is larger
compared to the other three decay modes (about 10\%). We therefore
allow only one \Ds\ candidate per event by choosing the
$\Dsm\mu^+$~combination with the largest probability from the combined
vertex fit (see Sec.~\ref{decaylength}).

The $K^+K^-$ invariant mass distribution for the $\Dsmu$ sample
is shown in Figure~\ref{ds_mass}d) with the fit result overlaid.
To obtain the number of $\phi$ signal events, we fit a second order
polynomial together with a Breit-Wigner
line shape 
convoluted with a Gaussian 
to account for detector resolution.
We find $205\pm38$ $\phi$ signal events and measure the $\phi$ mass to
be $(1020.1\pm0.5)$~\mevcc\ in agreement with the world average $\phi$
mass~\cite{PDG}.  
The shaded histogram in Fig.~\ref{ds_mass}d) 
shows the `wrong-sign' $KK$ mass spectrum, where we consider events with
same sign $K^{\pm}K^{\pm}$ or  $\mu^{\pm}\mu^{\pm}$ combinations as
`wrong-sign'. For display purposes the `wrong-sign' distribution is 
scaled by a factor of 0.6 to the same area as the combinatorial
background of the `right-sign' 
$K^+K^-$ distribution. The 
`wrong-sign' distribution describes very well the shape of the
combinatorial $K^+K^-$ background. No indication of a $\phi$ signal is
evident in the `wrong-sign' distribution.

\subsection{The Monte Carlo Simulation}
\label{mc}

Some quantities in this analysis like efficiencies or the $K$-factor
distribution further described in Sec.~\ref{decaylength} are
determined using 
a Monte Carlo (MC) calculation of $b$~quark production and $B$~meson decay
followed by a simulation of the detector response to the final state
particles. Since we extract only
kinematic quantities of the $B$~hadron decay from this Monte Carlo
study, we do not simulate the underlying event from the $p\bar p$
scattering nor include fragmentation products, but generate only
$B$~hadrons and their decay products.  

The MC simulation begins with a model of $b$~quark
production based on a next-to-leading order QCD
calculation~\cite{nloqcd}. This calculation employs the MRSD0 parton
distribution function~\cite{mrs} to model the kinematics of the
initial state partons. We generate $b$~quarks in the rapidity
interval $|y_b| < 1.0$ with a minimum $\Pt$ for the $b$~quark of
8~\gevc\ and 5~\gevc\ to simulate events corresponding to the single
lepton and dimuon data samples, respectively. These $\Pt$ requirements are
chosen in a way to avoid any biases in the $B$~meson kinematic
distributions after the application of the kinematic cuts used in the
analysis. The $b$~quarks are fragmented into $B$~mesons according to
a model using the Peterson fragmentation function~\cite{Peterson} with
a Peterson parameter of $\epsilon_b = 0.006$.
The bottom and charm hadrons are decayed into the various final
states using branching ratios and decay kinematics governed by the
world average masses and 
lifetimes of the involved particles~\cite{PDG}. 

Events with a lepton above a momentum threshold corresponding to the
appropriate hardware trigger are kept based on
an efficiency parameterization of the CFT trigger that depends on the
lepton $\Pt$. 
The accepted events are passed through a simulation
of the CDF detector 
that is based on parameterizations and simple models
of the detector response, which are functions of the particle
kinematics.
After the simulation of the CDF detector, the same selection criteria
applied to the data are imposed on the Monte Carlo events.

\section{The \Bs\ Lifetime Analysis}
\label{lifeana}

In this Section, we describe the measurement of the \Bs\ lifetime
starting with the determination of the primary event vertex followed
by the reconstruction of the \Bs\ decay length. In order to
determine the number of \Ds\ signal events used as a constraint in the
lifetime fit, a reflection from $D^-$ decays has to be considered for
the $\Dsm \ra \kstark$ and $\Dsm \ra \ksk$ decay modes. In Section~\ref{nonbsbg}
background from non-\Bs\ decays is discussed, while the lifetime
fit is detailed in Sec.~\ref{fitter}. The \Bs\ lifetime fit results are
then presented together with the determination of the systematic
uncertainties.  

\subsection{The Primary Event Vertex}
\label{primvx}

The \Bs\ lifetime reported in this paper is based on measuring the
distance between the primary $p\bar p$ event vertex and the secondary
\Bs\ decay vertex in the transverse plane. 
We first identify the $z$-position of the primary interaction vertex
using the tracks reconstructed in the VTX detector. 
These tracks, when
projected back to the beam axis, determine the longitudinal
location of the primary interaction with an accuracy of about
0.2~cm along the beam direction.
The primary vertices are distributed along the beam direction
according to a Gaussian function with a 
width of $\sim 30$~cm. 
On average during Run~I, the number of reconstructed interaction
vertices in a given event follows a Poisson distribution with a mean
of about 2.5.  
For the \Bs\ lifetime measurement, we determine the $z$-location of
the primary event vertex by choosing the 
$p\bar p$ interaction vertex recorded by the VTX which is closest to
the intercept of the 
lepton from the semileptonic \Bs\ decay with the beamline. We also
require the $z$-coordinates of all tracks from the \Ds\ decay to be
within 5~cm of the $z$-coordinate of this primary vertex. 

The transverse position of the primary event vertex is determined by using  
the average beam position through the detector together with the
knowledge of the longitudinal primary vertex position from the VTX.
The average beam position is calculated offline for each
data acquisition run. This calculation yields 
a transverse profile of the Tevatron beam which is circular with an RMS  
of $\sim 25~\mu$m in both the $x$- and $y$-directions.
We find that the average beam trajectory is stable over the period that a given
$p\bar p$~beam is stored in the Tevatron Collider. 
A detailed description of the determination of the average beamline
can be found in Ref.~\cite{lifetimeprd}. For the \Bs\ lifetime measurement,
we consider only events from data runs with a sufficiently large number of
collected events to allow 
a good determination of the run averaged beamline.
In this analysis, we choose not to measure the primary
vertex event-by-event because the presence of a second $b$ quark decay
in the event coupled with the low multiplicity in semileptonic
$B$~decays can lead to a systematic bias in the lifetime determination. 

\subsection{The Decay Length Reconstruction}
\label{decaylength}

The tracks forming the \Ds\ candidate are 
refit with a common vertex constraint referred to as the tertiary vertex
$V_{\Dsm}$. 
The secondary vertex where the \Bs\ decays to a lepton and a \Ds\
(referred to as $V_{\Bsm}$) is obtained by simultaneously 
intersecting the trajectory of the lepton track with the flight path 
of the \Ds\ candidate. 
Since we fully reconstruct the \Ds~meson in the $\phipi$, $\kstark$, and
$\ksk$ decay modes, we know the \Ds\ flight path. In the 
$\Dsm \ra \phil$ channel, where we do not fully reconstruct the
\Ds~meson, we use the $\phi\mu^-$ flight direction as a good estimate
of the \Ds\ flight path.

The confidence level of the combined vertex fit  
is required to be greater than 1\%.
Furthermore, we require that the 
reconstructed \Ds\ decay vertex $V_{\Dsm}$ be positively displaced 
from the primary vertex as projected along the direction of the \Ds\ 
momentum. 

\begin{figure}[tbp]\centering
\begin{picture}(160,85)(0,-5)
\centerline{
\put(20,67){\large\bf (a)}
\put(105,67){\large\bf (b)}
\put(30,0){\bf \boldmath{$\Pt(\Dsl)/\Pt(\Bsm)$}}
\put(115,0){\bf \boldmath{$\Pt(\Dsl)/\Pt(\Bsm)$}}
\epsfysize=8.5cm
\epsffile{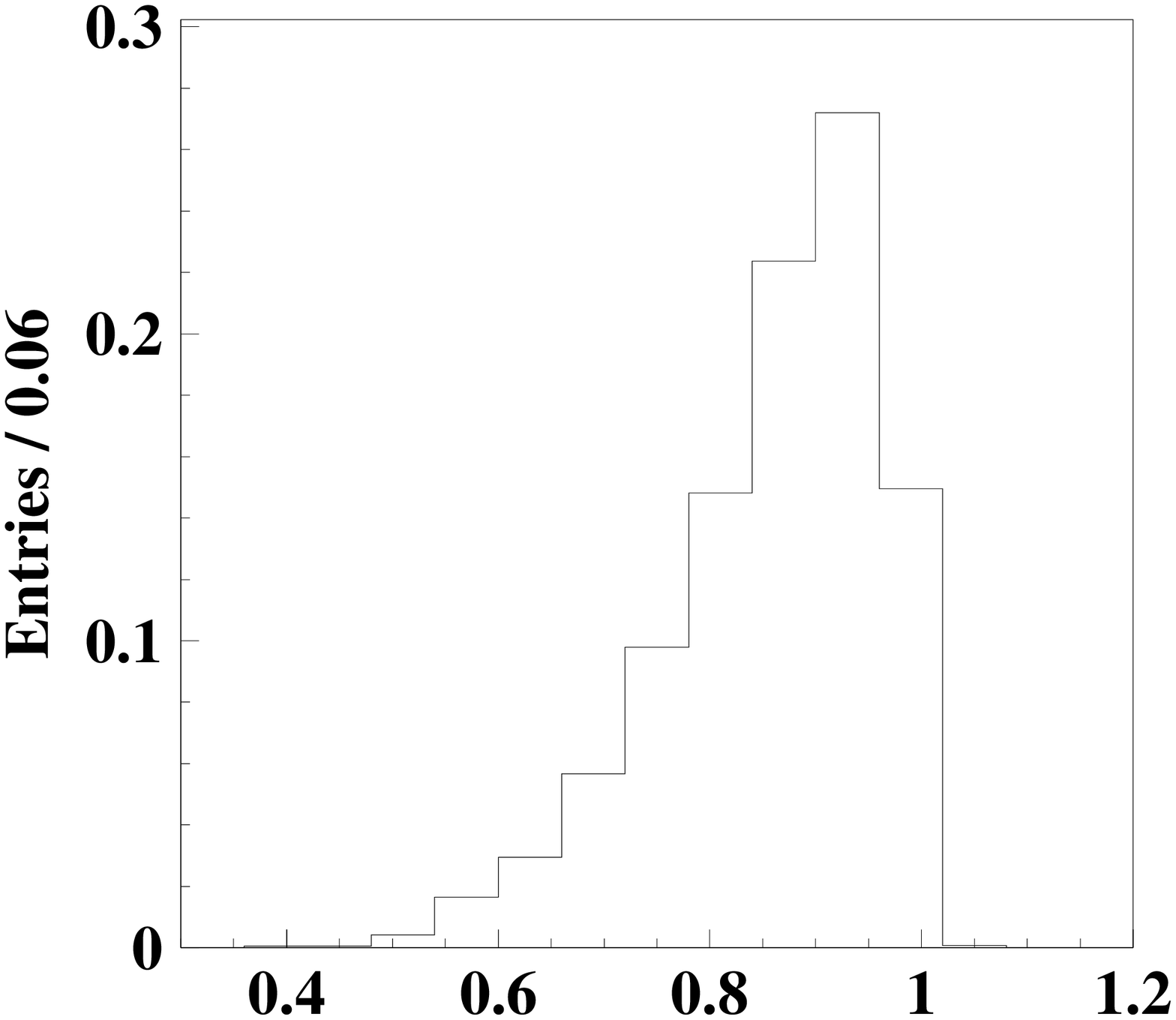}
\hspace*{-0.5cm}
\epsfysize=8.5cm
\epsffile{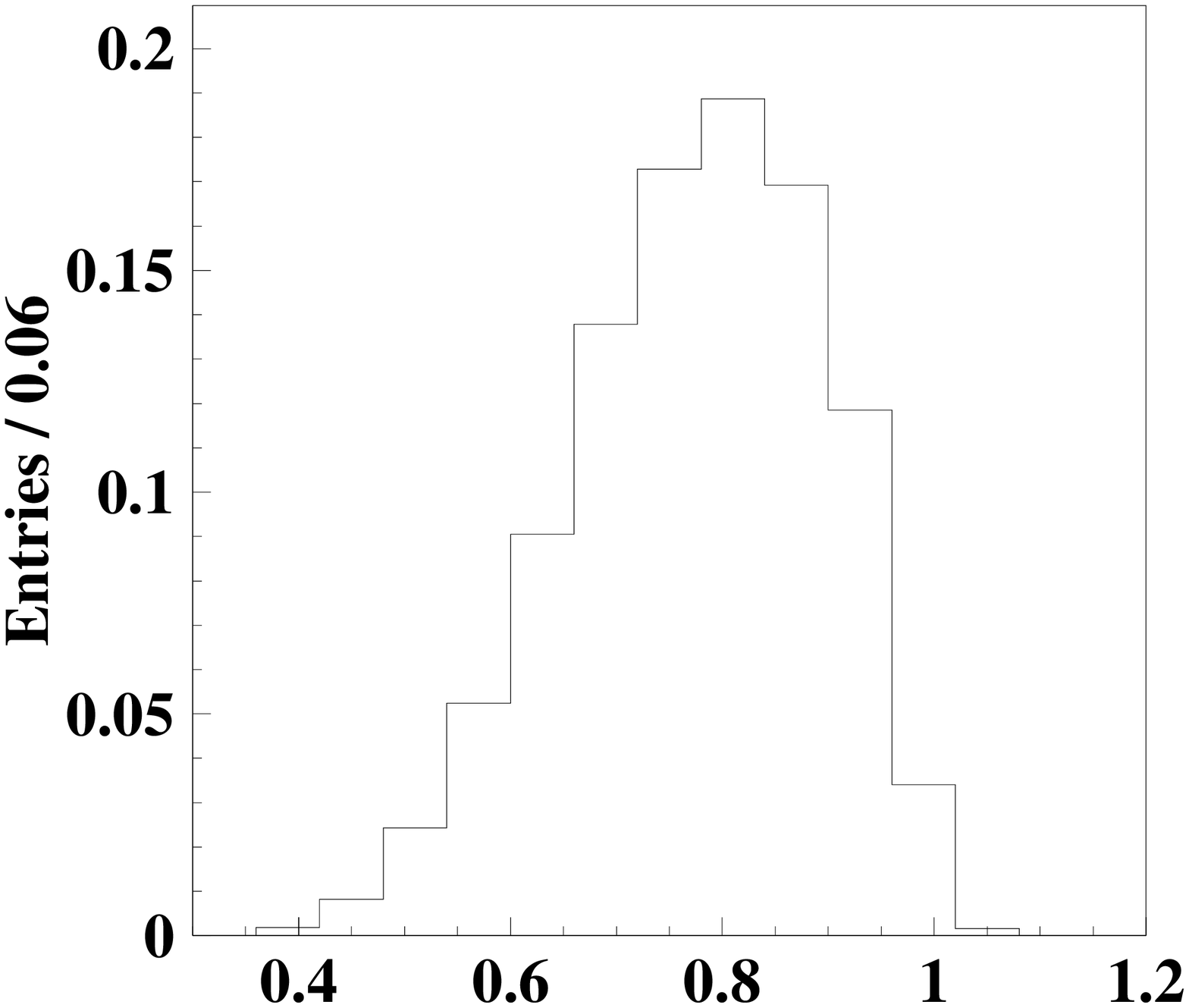}
}
\end{picture}
\caption{Normalized $K$-factor distributions $\Pt(\Dsl)/\Pt(\Bsm)$, for 
$\Bsm \ra \Dsl \nu X$ Monte Carlo decays with (a) $\Dsm \ra \phipi$ and
(b) $\Dsm \ra \phil$.}
\label{kdist}
\end{figure}

The transverse decay length $\lxy(\Bsm)$ is defined as the displacement 
$\vec X$ in the transverse plane of 
$V_{\Bsm}$ from the primary event vertex projected onto the $\Dsl$ momentum:
\begin {equation}
\lxy(\Bsm) = \frac{\vec X \cdot \vec{\Pt}(\Dsl)}{|\vec{\Pt}(\Dsl)|}.
\end{equation}
$\lxy$ is a signed variable which can be negative for the
configuration where the particle seems to decay before the point where
it is produced. 
The \Bs\ meson decay time is given by
\begin {equation}
c\,t\,(\Bsm) = \lxy \, \frac{{\rm m}(\Bsm)}{\Pt(\Bsm)}, 
\end{equation}
where m$(\Bsm)$ is the \Bs\ mass~\cite{PDG}.
Since we do not fully reconstruct the \Bs\ meson in our analysis, we 
define the `pseudo-proper decay length'
\begin {equation}
\lambda = \lxy \, \frac{{\rm m}(\Bsm)}{\Pt(\Dsl)},
\end{equation}
which has a typical uncertainty of $\sim 60~\mu$m including the
contribution from the finite size of the primary event vertex. 
In addition, we introduce a correction factor
\begin {equation}
K = \frac{\Pt(\Dsl)}{\Pt(\Bsm)},
\end{equation}
to correct between the reconstructed $\Pt(\Dsl)$ and the unknown $\Pt(\Bsm)$ in
the data. The \Bs\ meson decay time is then given as
\begin {equation}
c\,t\,(\Bsm) = \lxy \, \frac{{\rm m}(\Bsm)}{\Pt(\Dsl)} \times K.
\end{equation}
The correction between $\Pt(\Dsl)$ and $\Pt(\Bsm)$
is done statistically 
by smearing an exponential decay distribution with a Monte Carlo
distribution of the correction factor $K$  
when extracting $c\tau(\Bsm)$
from the pseudo-proper decay length in the lifetime fit as described in
Sec.~\ref{fitter}. 
The $K$-distribution is obtained from $\Dsl$ combinations which
originate from a Monte Carlo simulation (see Sec.~\ref{mc})
of semileptonic \Bs~decays into 
$D_{\mbox{\sf s}}^{(*)-}\ell^+ X$
including $D_{\mbox{\sf s}}^{(*)-}\tau^+ X$ with $\tau^+ \ra \ell^+ X$.
As an example, the $K$-distribution 
is shown for the $\Dsm \ra \phipi$ and   
$\Dsm \ra \phil$ decay modes in Figures~\ref{kdist}a) and \ref{kdist}b),
respectively. 
The $K$-distributions have mean values of 0.86 and 0.77 with RMS
values of 0.10 and
0.12 for the $\Dsm \ra \phipi$ and  
$\Dsm \ra \phil$ modes, respectively.
The $K$-distribution is 
approximately constant as a function of $\Pt(\Dsl)$ 
for the range of $\Pt(\Dsl)$ corresponding to our data.  

To ensure a precise \Bs~lifetime determination, we consider only \Bs\
candidates for which the pseudo-proper decay length is measured with an
uncertainty of less than 0.1~cm.  
We also require that the \Ds\ candidates have a proper decay length
measured between $V_{\Bsm}$ and $V_{\Dsm}$ 
of less than 0.1~cm and that the uncertainty on this proper decay length 
is less than 0.1~cm.  This requirement removes
background events with very long-lived \Ds\ candidates, where the long
extrapolation 
back to the \Bs\ decay vertex results in a poor vertex measurement.
These requirements have already been applied to the \Ds\ mass
distributions shown in Fig.~\ref{ds_mass}.

\subsection{The Determination of the Reflection from $D^-$}
\label{reflect}

\begin{figure}[tbp]\centering
\begin{picture}(160,180)(5,-5)
\centerline{
\put(23,105){\large\bf (a)}
\put(23,71){\large\bf (c)}
\put(108,105){\large\bf (b)}
\put(108,71){\large\bf (d)}
\epsfysize=16.5cm
\epsffile[5 10 520 520]{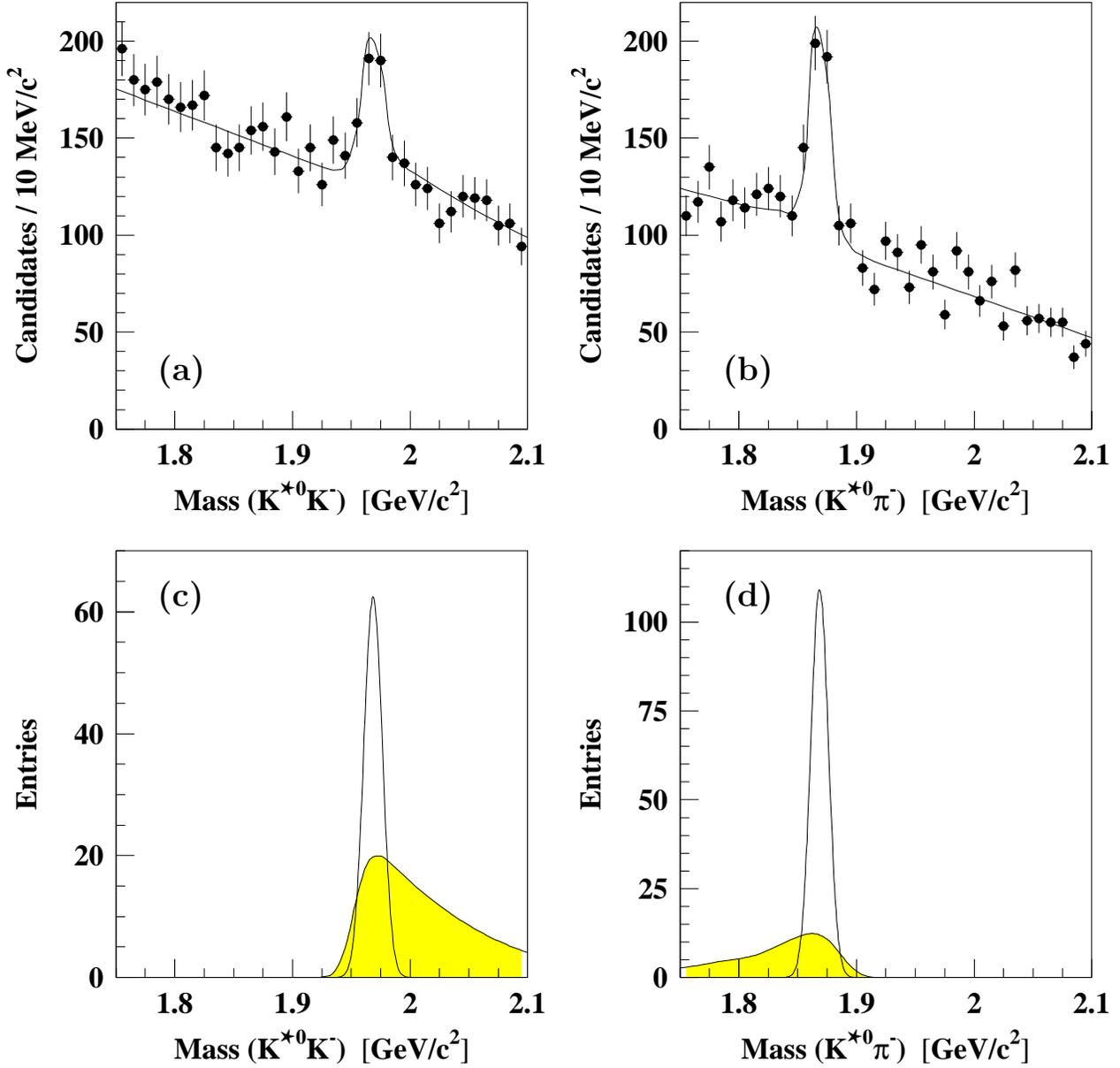}
}
\end{picture}
\caption{(a) Mass distributions for candidates in the $\Dsm \ra \kstark$ 
decay mode. (b) Mass distribution if these candidates are assumed to be
$D^- \ra \kstarpi$. 
(c) Distribution of the $\Dsm \ra \kstark$ signal
and the reflection from $D^- \ra \kstarpi$ (shaded
area) as obtained from Monte Carlo simulations. Normalizations are determined
from the simultaneous fit described in the text. 
(d) Mass distribution of the corresponding $D^- \ra \kstarpi$ signal
and the reflection from $\Dsm \ra \kstark$ (shaded area).}
\label{dreflect}
\end{figure}

The reconstructions of the \Ds\ decay modes into $\kstark$ and $\ksk$
suffer from reflections 
of $D^- \ra \kstarpi$ and $D^- \ra K^0_S \pi^-$, respectively, where the
$\pi^-$ is incorrectly assigned the kaon mass. We will discuss this
reflection from $D^-$ and the determination of the true number of events from
the \Ds\ decay with the example of the $\Dsm \ra \kstark$ mode.
The effect of this $K$-$\pi$ misassignment can be seen 
in Figure~\ref{dreflect}; events from a $B \ra D^- \ell
\nu X$ Monte Carlo simulation with $D^- \ra \kstarpi$ yield an invariant mass
distribution 
indicated by the shape of the shaded area in Fig.~\ref{dreflect}c) if
they are reconstructed as $\Bsm \ra \Dsm \ell \nu X$ with $\Dsm \ra
\kstark$, misinterpreting the $\pi^-$ as $K^-$. A significant portion
of this $D^-$ reflection lies at the \Ds\ mass peak.

Although we have already tightened our d$E$/d$x$ likelihood ratio to
better identify the $K^-$ track as a kaon, CDF's d$E$/d$x$
capabilities with a $\pi/K$ separation of about $1 \sigma$ for tracks 
with $\Pt$ greater than about 1~\gevc\ are not sufficient to remove
this $D^-$ reflection. Applying a $D^-$ mass
veto by rejecting all $\kstark$ combinations which are within a $\pm 3
\sigma$ window around the nominal $D^-$ mass when
reconstructed as $\kstarpi$, distorts the $\kstark$ mass
distribution. It would be very difficult to estimate the remaining
\Ds\ signal from that distribution and use it as input to the \Bs\
lifetime fit. We 
therefore choose to measure the $D^-$ reflection directly from our
data and account for the $D^-$ component in the \Bs\ lifetime fit. We
use two methods to determine the $D^-$ reflection in our
data.

The first method performs a simultaneous fit to the $\kstark$ and
$\kstarpi$ invariant mass distributions, where the $\kstarpi$ mass
distribution is created by switching the mass assignment on the
$K^-$ track to be a pion. Figure~\ref{dreflect}a) shows the $\kstark$
mass distribution, while the corresponding $\kstarpi$ mass distribution
is displayed 
in Fig.~\ref{dreflect}b). Each distribution is described by a Gaussian
for the corresponding $D^-$ and \Ds\ signal as shown in
Figures~\ref{dreflect}c) and \ref{dreflect}d) plus a linear lineshape
to parameterize the combinatorial background. The shape of the
corresponding $D^-$ or \Ds\ reflection as obtained from a Monte Carlo   
simulation is also included in the fit as displayed in
Fig.~\ref{dreflect}c) and d) as the shaded areas. The two mass
distributions are fit
simultaneously with the number of events in the Gaussian \Ds\ ($D^-$)
signal constrained to the number of events in the corresponding $D^-$
(\Ds) reflection. In addition, the difference between the \Ds\ and $D^-$
mass values is fixed based on their nominal mass values~\cite{PDG}.
The fit returns $123\pm25$ \Ds\ signal events and $80\pm10$ events from
the $D^-$ reflection within the \Ds\ signal region defined 
in Sec.~\ref{fitter}. The fit result is shown in Figure~\ref{dreflect}a)--d). 
We perform studies using MC pseudo-experiments to
verify the validity of this method~\cite{thesis}. We find that the
simultaneous 
fitting method returns the number of true \Ds~events in our MC
studies with no bias and the error obtained from the fit to the data
agrees with the expected uncertainty of this technique for our sample size.

The second method for determining the amount of $D^-$ reflection in our
\Ds\ signal events exploits the difference between the $D^-$~lifetime
($\tau(D^-) = (1.057\pm0.0015)$~ps~\cite{PDG}) 
and the \Ds~lifetime
($\tau(\Dsm) = (0.467\pm0.0017)$~ps~\cite{PDG}). As described in
Sec.~\ref{fitresults}, we can determine the \Ds~lifetime in our fit
for the \Bs\ lifetime. We modify the fitting method 
used to determine the \Ds\ lifetime 
in the following way: We replace the
exponential describing the \Ds\ signal by the sum of two exponentials,
one with the \Ds~lifetime and one with the $D^-$~lifetime 
(see Sec.~\ref{fitter} about the fitting method and 
Sec.~\ref{dgamogam} for an example of a two-lifetime fit).
We fix the \Ds\ and $D^-$ lifetimes to their nominal values~\cite{PDG} and 
allow the relative fractions of \Ds\ and $D^-$ to float in the
fit. With this method we obtain $129^{+31}_{-34}$ \Ds\ events and 
$84^{+34}_{-31}$ events attributed to the $D^-$ reflection. We again perform
studies using MC pseudo-experiments and verify the
validity of this method to work without any bias~\cite{thesis}. 

We determine the weighted average of \Ds\ events from both methods and
obtain a \Ds\ signal of $125\pm20$ events for the $\Dsm \ra \kstark$
decay. Both methods are also used to calculate the number of \Ds\ events and
the contribution from the $D^-$ reflection in the $\Dsm \ra
\ksk$ decay mode. We obtain
$33\pm8$ \Ds~signal events for the $\ksk$ mode. These numbers are displayed in
Table~\ref{numds} together with the numbers of \Ds\ signal events for
the $\Dsm \ra \phipi$ and $\Dsm \ra \phil$ decay modes.
As further described in Sec.~\ref{fitter}, these
event numbers are used as a constraint in the \Bs\ lifetime fit.
 
\subsection{The Non-\Bs\ Background}
\label{nonbsbg}

There are two possible sources of non-strange $B$~meson decays which can
lead to `right-sign' $\Dsl$ combinations.
The first process originates from the decays 
$\bar B^0 \ra D_{\mbox{\sf s}}^{(*)-}D^{(*)+}X$ and 
$B^- \ra D_{\mbox{\sf s}}^{(*)-}D^{(*)0}X$, with the $D^0$ or $D^+$
decaying semileptonically. These decays 
produce softer and less isolated leptons than the leptons from \Bs\
semileptonic  
decays. Therefore we expect the acceptance for this background source
relative to the \Bs\ signal to be quite small.
We use a Monte Carlo simulation of these events and estimate their
contribution $f_{\dsd}$ in the following way:
\begin {equation}
f_{\dsd} = \epsilon_{\rm rel} \cdot \frac{f_u + f_d}{\fs} \cdot
\frac{BR(B \ra D_{\mbox{\sf s}}^{(*)}D^{(*)}X)}
{BR(\Bsm \ra D_{\mbox{\sf s}}^{(*)}\ell \nu X)}.
\end {equation}
We use the following branching ratios and fragmentation fractions from
the Particle Data Group~\cite{PDG}: 
$BR(\Bsm \ra D_{\mbox{\sf s}}^{(*)}\ell \nu X) = (7.6\pm2.4)\%$,
$BR(B \ra D_{\mbox{\sf s}}^{(*)}D^{(*)}X) = (4.9\pm1.1)\%$,
$f_u = f_d = (37.8\pm2.2)\%$, and $\fs = (11.2\pm2.2)\%$. 
$\epsilon_{\rm rel}$ is the ratio of efficiencies and acceptances for
both decays obtained from a Monte Carlo simulation:
\begin{equation}
\epsilon_{\rm rel} =
\frac{\epsilon\,(B \ra D_{\mbox{\sf s}}^{(*)}D^{(*)}X)}
{\epsilon\,(\Bsm \ra D_{\mbox{\sf s}}^{(*)}\ell \nu X)}.
\end{equation}
The values obtained for $\epsilon_{\rm rel}$ are in the order of 0.5\%
to 1\%. 
The calculated fractions $f_{\dsd}$ for each \Ds\ decay mode are
compiled in Table~\ref{numds}. The $f_{\dsd}$ fraction is larger for
the $\Dsm \ra \phil$ decay mode because of the on average softer
$B$~meson 
momentum in the dimuon data sample compared to the single lepton trigger
events.

The second process is a four 
body decay $B^0/B^+ \rightarrow \Dsm {\cal K} \ell^+ {\nu} X$, where 
${\cal K}$ denotes any type of strange meson. Because of the low probability
of producing $s\bar{s}$ pairs and the limited phase space, this process 
is suppressed and has not been observed experimentally~\cite{exp_limit}.
Based on the quoted limit 
${\rm BR}(B^0/B^+ \rightarrow \Dsm \ell^+ \nu X) < 0.9\,\%$
(90\% CL)~\cite{PDG,exp_limit} 
and our detection efficiency determined from MC simulation, we expect less
than 1.0\,\% of our $\Dsl$ combinations to originate from this source. 

We also consider events from
$\Bsm \ra D_{\mbox{\sf s}}^{(*)+}D_{\mbox{\sf s}}^{(*)-} X$ decays, with one 
$D_{\mbox{\sf s}}$ decaying semileptonically.
This contribution to our \Bs\ lifetime sample is
determined from Monte Carlo studies in the same way as described above
for the $B \ra D_{\mbox{\sf s}}^{(*)}D^{(*)}X$ background.
The obtained fractions $f_{\dsds}$ from these decays are small and 
compiled in Table~\ref{numds}. Finally, backgrounds with a real
\Ds~meson and a fake lepton from decays such as 
$B \ra D_{\mbox{\sf s}}^{(*)}D^{(*)} X$ with a hadron from the
$D^{(*)}$ decay faking a lepton are negligible due to the low
probability of a hadron faking a lepton.

In summary, the contribution of all above physics backgrounds is quite small 
compared to the combinatorial background. We account for contributions from 
$B \ra D_{\mbox{\sf s}}^{(*)}D^{(*)}X$ and 
$\Bsm \ra D_{\mbox{\sf s}}^{(*)+}D_{\mbox{\sf s}}^{(*)-} X$
decays in our lifetime
fit described next, 
and treat the contribution of $B^0/B^+ \ra \Dsm {\cal K} \ell^+ \nu X $ decays
as a source of systematic uncertainty in the \Bs\ lifetime measurement. 

\subsection{The Description of the Lifetime Fit}
\label{fitter}

As input to the \Bs\ lifetime fit, we define a signal
sample using a \Ds\ mass window from 1.944~\gevcc\ to 1.994~\gevcc\  
for the $\Dsm \ra \phipi$, $\kstark$, and $\ksk$ decay modes and a $\phi$
signal window from 1.0094~\gevcc\ to 1.0294~\gevcc\ for the 
$\Dsm \ra \phil$ decay channel.
The numbers of events in the signal samples can be found for the four
decay modes in Table~\ref{numds}. 
To model the pseudo-proper decay length distribution of the combinatorial
background events  
contained in the signal sample, we define a background sample which 
consists of `right-sign' events from the \Ds\ sidebands
(1.884~\gevcc\ -- 1.934~\gevcc\
and 2.004~\gevcc\ -- 2.054~\gevcc) and `wrong-sign' events 
from the interval 1.884~\gevcc\ to 2.054~\gevcc. 
For the $\Dsm \ra \phil$ decay mode the $\phi$ sidebands are defined from 
0.9844~\gevcc\ to 1.0044~\gevcc\ and from 1.0344~\gevcc\ to 1.0544~\gevcc, 
while the `wrong-sign' combinations are taken from the region 
0.9844~\gevcc\  to 1.0544~\gevcc.
We assume the combinatorial background to originate from random track
combinations and therefore use the sidebands to model the background
in the signal sample. This assumption is supported by the mass
distribution of the `wrong-sign'
combinations where no enhancement is visible at the \Ds~mass.
By adding the `wrong-sign' combinations to the `right-sign' sideband
events, we better constrain the shape of the combinatorial
background events in the \Ds\ signal samples for decay channels with
low combinatorial background like the $\Dsm \ra \phipi$ mode.  
 
The pseudo-proper decay length distribution obtained from the signal sample
is fit using an unbinned maximum
log-likelihood method. Both the \Bs\ lifetime, denoted as $c\tau$
below, and the background shape  
are determined in a simultaneous fit using the signal and background
samples. Thus the likelihood function ${\cal L}$ is a combination of
two parts:  
\begin{equation}
{\cal L}  =  \prod^{N_S}_i\, [f_{\rm sig}{\cal F}^i_{\rm sig}
 + (1 - f_{\rm sig}){\cal F}^i_{\rm bg}\, ] 
\cdot \prod^{N_B}_j{\cal F}^j_{\rm bg},
\end{equation}
where $N_S$ and $N_B$ are the number of events in the signal and background 
samples.
$f_{\rm sig}$ is the ratio of \Ds\ signal events 
obtained from the \Ds\ mass
distributions (see Table~\ref{numds}) to the total number of events
in the signal sample. To constrain $f_{\rm sig}$ we factor in an
additional $\chi^2$~term to the 
likelihood function $\cal L$ above to constrain the number of \Ds\
signal events obtained from the invariant mass distributions
within their uncertainty.

The signal probability function ${\cal F}_{\rm sig}$ consists of a normalized 
decay exponential function convoluted with 
a Gaussian resolution function ${\cal G}$ and is smeared with a normalized
$K$-distribution ${\cal H}(K)$:
\begin{equation}
{\cal F}^i_{\rm sig}(x) = 
\int dK\, {\cal H}(K)\, \left[
\frac{K}{c\tau} \exp\{-\frac{K x}{c\tau}\} 
\otimes {\cal G}(\lambda^i |\, x,{\rm s}\sigma^i)
\right].
\end{equation}
Here, $\lambda^i$ is the measured pseudo-proper decay length with uncertainty
$\sigma^i$ and 
$x$ is the true pseudo-proper decay length.
Because of systematic uncertainties in the overall scale of the decay
length uncertainties, which we estimate on an event-by-event basis, we
introduce a scale factor, s, which is a free parameter in the \Bs\
lifetime fit.  We subsequently vary s in the fits to determine the
sensitivity of our measurement to this uncertainty.
The integration over the momentum ratio $K$ is approximated by a
finite sum
\begin{equation}
\int dK\, {\cal H}(K) \ra
\sum_{i}\, \Delta K\,{\cal H}(K_i), 
\end{equation}
where the sum is taken over bin $i$ of a histogrammed distribution
${\cal H}(K_i)$ with bin width $\Delta K$ as shown e.g.~in Figure~\ref{kdist}.

The background probability function ${\cal F}_{\rm bg}$ is parameterized
by a Gaussian centered at zero,  
a negative exponential tail,
and a positive decay exponential to characterize the contribution of
heavy flavor decays in the background sample:
\begin{eqnarray}
{\cal F}^i_{\rm bg}(x) & = &
(1-f_{+}-f_{-}) \, {\cal G}(\lambda^i |\, x,{\rm s}\sigma^i) + \nonumber \\
& + & \frac{f_{+}}{\lambda_{+}} \exp\{-\frac{x}{\lambda_+}\} 
  \otimes {\cal G}(\lambda^i |\, x,{\rm s}\sigma^i) + \nonumber \\
& + & \frac{f_{-}}{\lambda_{-}} \exp\{-\frac{x}{\lambda_-}\} 
  \otimes {\cal G}(\lambda^i |\, x,{\rm s}\sigma^i).
\label{eq:bgparam}
\end{eqnarray}
Here, $f_{\pm}$ are the fractions of positive and negative lifetime
backgrounds and
$\lambda_{\pm}$ are the effective lifetimes of those backgrounds. 
We verify the parameters $f_{\pm}$ and $\lambda_{\pm}$ agree with the
`right-sign' sideband events and the `wrong-sign' combinations separately,
allowing us to combine both samples resulting in the background samples
described above.

The events originating from the $D^-$ reflection in the $\Dsm \ra \kstark$
and $\Dsm \ra \ksk$ decays (see Sec.~\ref{reflect}) are also accounted
for in the likelihood function by a term
\begin{equation}
\int dK\, {\cal H}(K)\, \left[
f_{D^-}\frac{K}{c\tau(B)} \exp\{-\frac{K x}{c\tau(B)}\} 
\otimes {\cal G}(\lambda^i |\, x,{\rm s}\sigma^i)
\right],
\end{equation}
where $f_{D^-}$ refers to the fraction of the $D^-$ reflection in the
\Ds\ sample and
$c\tau(B)$ is taken to be the world average $B^0$ lifetime~\cite{PDG}.

\subsection{The Fit Results}
\label{fitresults}

\begin{figure}[tbp]\centering
\begin{picture}(160,160)(5,-5)
\centerline{
\put(30,147){\Large\bf (a)}
\put(30,67){\Large\bf (b)}
\epsfysize=16cm
\epsffile[10 10 545 515]{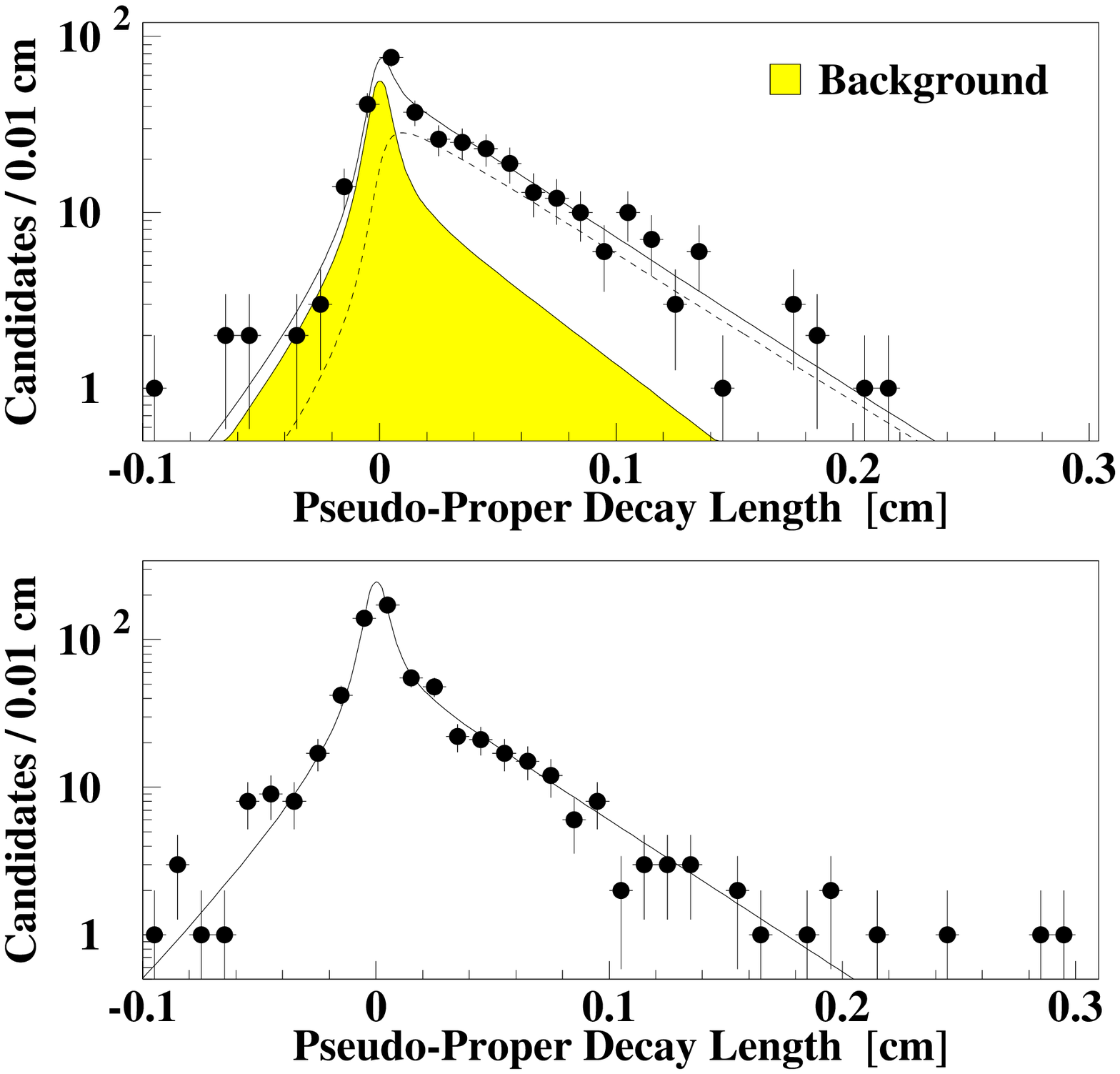}
}
\end{picture}
\caption{
(a) \Bs\ pseudo-proper decay length distribution for the $\Dsm \ra
\phipi$ signal 
sample with the result of the fit superimposed.
The dashed line is the \Bs\ signal contribution, while the
shaded curve represents the contribution from the combinatorial
background.
(b) Pseudo-proper decay length distribution
for the background sample with the fit result superimposed.}
\label{ctau_phipi}
\end{figure}

\begin{figure}[tbp]\centering
\begin{picture}(160,160)(5,-5)
\centerline{
\put(30,147){\Large\bf (a)}
\put(30,67){\Large\bf (b)}
\epsfysize=16cm
\epsffile[10 10 545 515]{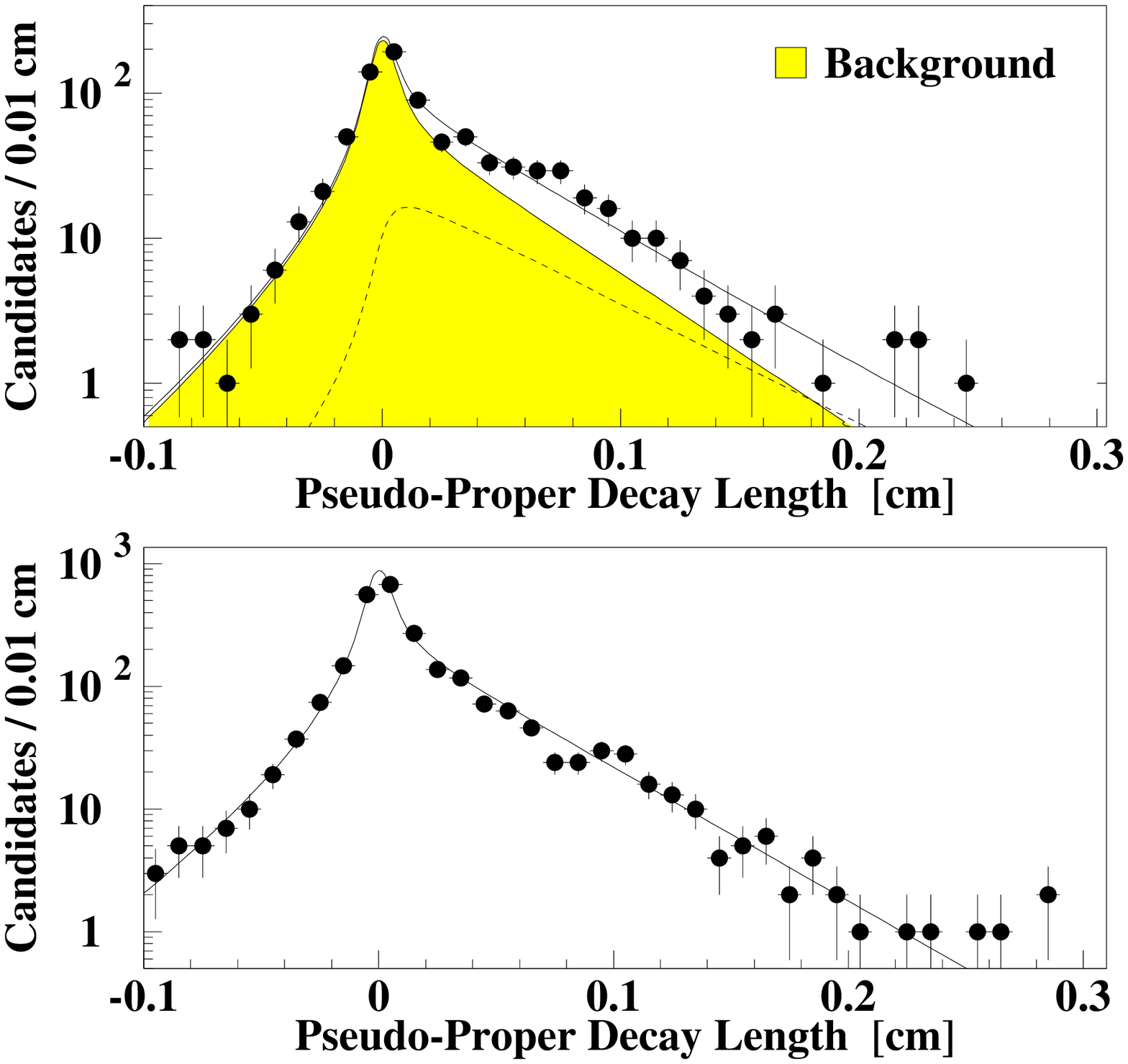}
}
\end{picture}
\caption{
(a) \Bs\ pseudo-proper decay length distribution for the $\Dsm \ra
\kstark$ signal 
sample with the result of the fit superimposed.
The dashed line is the \Bs\ signal contribution, while the
shaded curve represents the contribution from the combinatorial
background.
(b) Pseudo-proper decay length distribution
for the background sample with the fit result superimposed.}
\label{ctau_kstark}
\end{figure}

\begin{figure}[tbp]\centering
\begin{picture}(160,160)(5,-5)
\centerline{
\put(30,147){\Large\bf (a)}
\put(30,67){\Large\bf (b)}
\epsfysize=16cm
\epsffile[10 10 545 515]{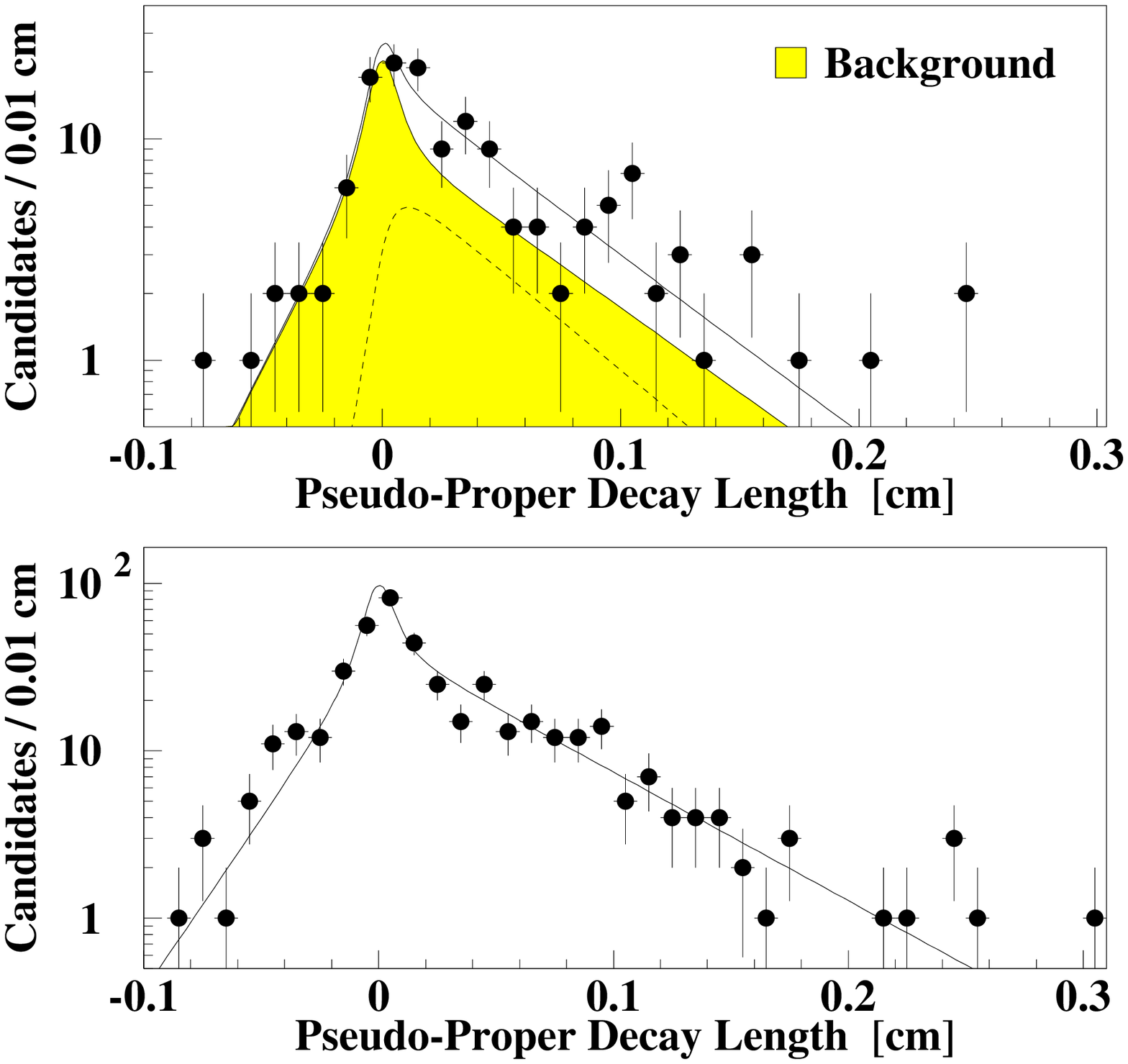}
}
\end{picture}
\caption{
(a) \Bs\ pseudo-proper decay length distribution for the $\Dsm \ra \ksk$ signal
sample with the result of the fit superimposed.
The dashed line is the \Bs\ signal contribution, while the
shaded curve represents the contribution from the combinatorial
background.
(b) Pseudo-proper decay length distribution
for the background sample with the fit result superimposed.}
\label{ctau_ksk}
\end{figure}

\begin{figure}[tbp]\centering
\begin{picture}(160,160)(5,-5)
\centerline{
\put(30,147){\Large\bf (a)}
\put(30,67){\Large\bf (b)}
\epsfysize=16cm
\epsffile[10 10 545 515]{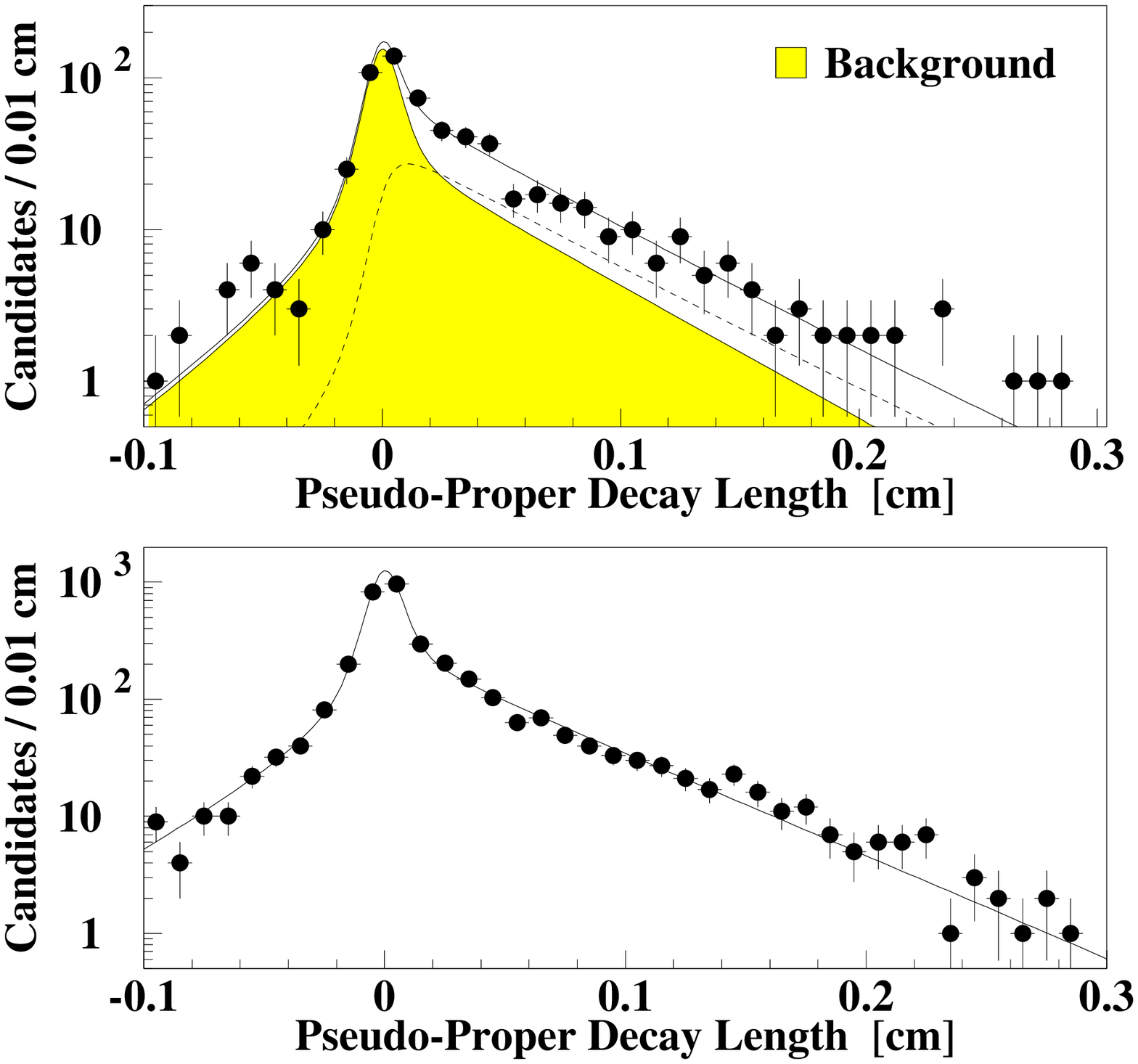}
}
\end{picture}
\caption{
(a) \Bs\ pseudo-proper decay length distribution for the $\Dsm \ra \phil$ signal
sample with the result of the fit superimposed.
The dashed line is the \Bs\ signal contribution, while the
shaded curve represents the contribution from the combinatorial
background.
(b) Pseudo-proper decay length distribution
for the background sample with the fit result superimposed.}
\label{ctau_phil}
\end{figure}

We first determine the \Bs\ lifetime for each of the four \Ds\ decay
channels individually. 
The parameters allowed to float in the fit are the \Bs~lifetime,
$f_{\rm sig}$, $\lambda_{\pm}$, $f_{\pm}$, and the overall scale
factor s.
The fitted values for $c\tau(\Bsm)$ and their statistical uncertainties are
shown in Table~\ref{numds}, and are in good statistical agreement.
The pseudo-proper decay length distribution 
of the signal sample with the result of the fit superimposed  
is shown in Figure~\ref{ctau_phipi}a) for the $\Dsm \ra \phipi$ decay mode.
The dashed line represents the \Bs\ signal contribution, while
the shaded curve shows the sum of the background probability function over
the events in the signal sample.
The same distribution of the background sample is displayed in 
Figure~\ref{ctau_phipi}b) with the result of the fit superimposed. 
Figures~\ref{ctau_kstark}, \ref{ctau_ksk}, and \ref{ctau_phil} show
the corresponding distributions for the $\Dsm \ra \kstark$, $\Dsm \ra \ksk$, 
and $\Dsm \ra \phil$ decay modes, respectively. The combined \Bs\
lifetime from all four \Ds\ decay modes is determined from a
simultaneous fit to be $c\tau(\Bsm) = (408\ _{-27}^{+28})\ \mu$m or 
$\tau(\Bsm) = (1.36\ \pm 0.09)$~ps, where the errors shown are
statistical only. 

\begin{figure}[tbp]\centering
\begin{picture}(160,160)(5,-5)
\centerline{
\put(30,147){\Large\bf (a)}
\put(30,67){\Large\bf (b)}
\epsfysize=16cm
\epsffile[10 10 545 515]{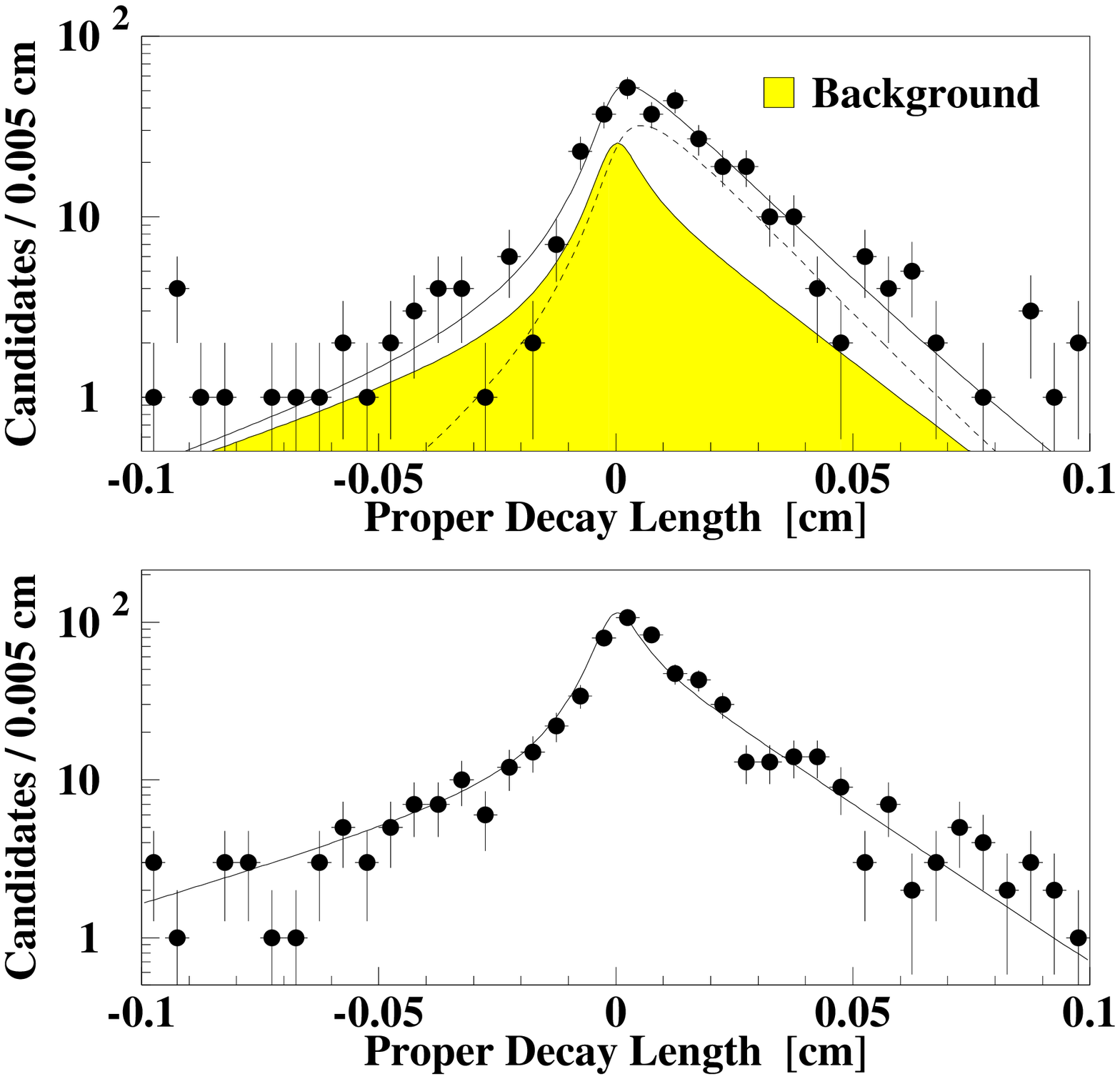}
}
\end{picture}
\caption{
(a) \Ds\ proper decay length distribution $c\tau(\Dsm)$ for the 
$\Dsm \ra \phipi$ signal sample with the result of the fit superimposed.
The dashed line is the \Ds\ signal contribution, while the
shaded curve represents the contribution from the combinatorial
background.
(b) Proper decay length distribution $c\tau(\Dsm)$
for the background sample with the fit result superimposed.}
\label{ctauds}
\end{figure}

As a consistency check, we use the $\Dsm \ra \phipi$ decay mode to also
fit the \Ds\ lifetime  from the  
proper decay length measured from the secondary vertex
$V_{B_{\mbox{\sf s}}}$ to the tertiary vertex $V_{D_{\mbox{\sf s}}}$.
Since the \Ds\ decay is fully reconstructed, its relativistic boost is known 
and a convolution with a $\Pt$-correction factor distribution in the fit does 
not apply.  The result is $c\tau(\Dsm) = (136\ _{-15}^{+17})\ \mu m$
(statistical error only),
which is consistent with the world average \Ds\ lifetime~\cite{PDG}.
Figures~\ref{ctauds}a) and ~\ref{ctauds}b) show the \Ds\ proper decay
length distributions for the
signal and background samples, respectively, with the results of the fit
superimposed.

\subsection{The Systematic Uncertainties}

Table~\ref{systematics} lists all sources of systematic uncertainty
considered in this analysis. 
The major contribution originates from the treatment of the background.
In particular, the following evaluations of systematic errors are
performed yielding the uncertainties reported in
Tab.~\ref{systematics}: 
\begin{itemize}
\item 	Background treatment: The combinatorial background in the
	signal sample is parameterized by the 
	positive and negative lifetimes
      	$\lambda_+$ and $\lambda_-$  
      	as well as their respective fractions $f_+$
      	and $f_-$ as described in Eq.~(\ref{eq:bgparam}). To evaluate
	the dependence of the \Bs~lifetime on the background
	parameterization, we vary $\lambda_{\pm}$ and $f_{\pm}$ within
	$\pm 1 \sigma$ of their values returned 
      	from the \Bs~lifetime fit.
	Since there is a correlation between the background parameters
	$\lambda_{\pm}$ and $f_{\pm}$ and the \Bs\ lifetime in the
	simultaneous 
	fit to the signal and background samples, part of this systematic
	uncertainty is already accounted for in the statistical error on
	$\tau(\Bsm)$. However, this correlation is small. 
	We therefore adopt this method as a conservative way to
	evaluate the systematic uncertainty from background treatment.
\item 	Non-\Bs\ backgrounds: These backgrounds (see Sec.~\ref{nonbsbg})
      	enter our fit as fixed fractions. 
      	We vary these fractions within $\pm 50\%$ of their reported values
      	(see Tab.~\ref{numds}) and repeat the fit. 
	We also evaluate the background from $B^0/B^+ \ra \Dsm
      	{\cal K} \ell^+ \nu X $ decays and consider its fraction to be
	2.0\%, twice the quoted limit~\cite{PDG,exp_limit}.
\item 	Decay length requirement: There are two requirements that can
	bias the \Bs\ 
       	lifetime result. These are the cut on $|c\tau(\Dsm)| < 0.1$~cm and the
       	requirement that the
       	reconstructed \Ds\ decay vertex $V_{\Dsm}$ be positively displaced 
	from the primary event vertex.
	To study the effect of these cuts, we use high statistics Monte
	Carlo samples. We first fit the lifetime with all
	the selection requirements, and then remove each cut individually
	noting the shift in the \Bs\ lifetime.
\item 	Momentum estimate: The \Bs\ lifetime result is sensitive to
	the distribution of the correction  
	factor $\cal{H}(K)$, 
	which can be affected by the lepton $\Pt$ cut and the
	decay kinematics.  For the standard fit we require 
	$\Pt(\mu) > 2.0$~\gevc\ for the $\Dsm \ra \phil$ decay mode and
	$\Pt(\ell) > 6.0$~\gevc\ for the other \Ds\ decay channels. To test
	the effect of the lepton $\Pt$ dependence, we generate new
	$K$-distributions for lower and higher lepton $\Pt$ cut values.  
	We also compare the effect on the kinematics of semileptonic
	\Bs\ decays using a pure 
	V-A decay versus semileptonic decays using the ISGW form
	factor~\cite{isgw}.
	In addition, an alternative $\Pt$ spectral shape of 
	$b$~quark production is considered based on a comparison of
	the lepton $\Pt$ shape in the data and in Monte Carlo
	events. Finally, the $K$-distribution is somewhat dependent on
	the electron identification. We study a possible
	incompleteness in the treatment of the electron selection with
	our Monte Carlo simulation and assign a systematic error of
	$\pm3\ \mu$m.
\item   Decay length resolution: Our uncertainty on the estimate of
	the decay length 
	resolution is expressed in the scale factor~s, which is fitted to 
	$1.29 \pm 0.03$. We fix the 
	scale factor at 1.0, and again at 1.38, the latter
	corresponding to a $+ 3 \sigma$ upward shift from the 
	fitted value, and repeat the \Bs\ lifetime fit.   
\item  	$D^-$ reflection: The reflection from $D^-$ in the 
	$\Dsm \ra \kstark$ and $\Dsm \ra \ksk$ decay modes changes the
	number of \Ds\ signal events in these two channels. We study
	the influence of the $D^-$ reflection by varying the number of
	\Ds\ signal events within their error as determined in
	Sec.~\ref{reflect}. 
\item   Detector alignment: We also account for a possible residual
	misalignment of the SVX and assign an error of $\pm 2\ \mu$m as
	further detailed in Ref.~\cite{lifetimeprd}.      	    	
\end{itemize}

The systematic uncertainties noted above have been combined in
quadrature. Quoting the statistical and systematic uncertainties
separately, we measure the \Bs\ lifetime using semileptonic \Bs~decays to be 
\begin {equation}
  \tau(\Bsm) = (1.36\ \pm 0.09 \ ^{+0.06}_{-0.05})\  {\rm ps},
\end {equation}
where the first error is statistical and the second systematic.
This result is currently the world's best measurement of the \Bs\ lifetime
from a single experiment. In comparison, the world average
\Bs~lifetime is $(1.57 \pm 0.08)$~ps~\cite{PDG}. This measurement
supersedes CDF's previously published \Bs~lifetime result 
of $\tau(\Bsm) = (1.42\,^{+0.27}_{-0.23}\pm0.11)$~ps using a data sample
corresponding to an integrated luminosity of
20~pb$^{-1}$~\cite{old_bs_cdf}.
 
Using the CDF average $B^0$ lifetime 
$\tau(B^0) = (1.513\pm0.053)$~ps~\cite{fumiprd},
we determine the $\Bsm/B^0$ lifetime ratio to be $0.899\pm0.072$ taking
correlated systematic uncertainties into account. 
However, ignoring the correlated systematic uncertainties increases
the error on the $\Bsm/B^0$ lifetime 
ratio only to $\pm 0.077$, since $\tau(\Bsm)/\tau(B^0)$ is dominated by the
statistical error on the \Bs\ lifetime measurement.

\section{The Determination of $\dgog$}
\label{dgamogam}

We examine the \Bs~meson pseudo-proper decay length distribution from $\Dsl$
correlations for a lifetime difference $\dgog$ between the two mass
eigenstates of 
the \Bs\ meson, $\Bsh$ and $\Bsl$. 
In the case of a lifetime difference in the \Bs\ system, the decay
length distribution for events from the semileptonic 
decay $\Bsm \ra \Dsm \ell^+ \nu X$
is expected to be governed by the sum of two exponentials.
We expand the likelihood fit to 
describe the \Bs\ pseudo-proper decay length distribution 
to a function of the form
\begin {equation}
{\cal F}(t) =
{\rm e}^{-\Gamma_H t} + {\rm e}^{-\Gamma_L t}\ \ \ \ \
{\rm with}\ \ \Gamma_{L,H} = \Gamma \pm \frac{\dgam}{2} 
= \Gamma \cdot (1 \pm \frac{1}{2} \frac{\dgam}{\Gamma}),
\label{eq:twolif}
\end{equation}
rather than fitting for just one exponential lifetime ${\rm e}^{-\Gamma t}$. 
The parameter $\dgog$ is the parameter we fit for. 
Since $\dgog$ is symmetric about zero, it is required to be positive.
In the case of a lifetime difference $\dgam \neq 0$, the total decay width
$\Gamma = 1/2 \cdot (\Gamma_H + \Gamma_L)$ and the mean \Bs\ lifetime 
$\tau_{\rm m}(\Bsm)$ obtained from a fit assuming a single \Bs\
lifetime, are no longer reciprocal to each other but follow the relation
\begin{equation}
\tau_{\rm m}(\Bsm) = \frac{1}{\Gamma} \cdot
         \frac{1+(\frac{\dgam}{2\,\Gamma})^2}
	      {1-(\frac{\dgam}{2\,\Gamma})^2}.
\label{eq:gamtau}
\end{equation}
We incorporate the relation in Eq.~(\ref{eq:gamtau}) into our
likelihood fitting function. We follow the suggestion given in
Ref.~\cite{bbd_bs} and fix the mean \Bs\ lifetime to the world average
$B^0$ lifetime since both lifetimes are expected to agree within
1\%~\cite{bigi,neubert}. This theoretical assumption can be verified
by the current world 
average $\tau(B^0) = 1.55 \pm 0.05$~ps and $\tau(\Bsm) =
1.57\pm0.08$~ps. The CDF average \Bs/$B^0$ lifetime ratio derived
above also supports this assumption.  

The fit returns 
$\dgog = 0.34\ ^{+0.31}_{-0.34}$, where the given error is 
statistical only. This indicates that with the current statistics of our $\Dsl$
sample we are not sensitive to a \Bs\ lifetime difference. Based on this fit
result, we integrate the normalized likelihood as a function of $\dgog$ and
find the 95\% confidence level (CL) limit at
\begin {equation}
  \frac{\dgam}{\Gamma} < 0.83\ \ \ (95\%\ {\rm CL}).
\end {equation}
This is the first experimental result for the lifetime difference in
the \Bs\ system.

Using a value of
$\dgam/\dm = (5.6\pm2.6)\cdot 10^{-3}$
from Ref.~\cite{bbd_bs} and setting $\tau_{\rm m}(\Bsm)$ to the world
average $B^0$~lifetime~\cite{PDG}, 
an upper limit on the \Bs\ mixing frequency
of $\dms < 96~{\rm ps}^{-1}$ (95\%~CL) can be determined within the
Standard Model. Including the
dependence on $\dgam/\dm$ and $\tau_{\rm m}(\Bsm)$ in our limit, we obtain
\begin {equation}
  \dms < 96~{\rm ps}^{-1} 
\times \left(\frac{5.6\cdot 10^{-3}}{\dgam/\dm}\right)
\times \left(\frac{1.55~{\rm ps}}{\tau_{\rm m}(\Bsm)}\right)
 \ \ \ (95\%\ {\rm CL}).
\end {equation}

\section{Conclusion}
\label{conc}

We have presented a measurement of the \Bs\ meson lifetime using
semileptonic \Bs~decays, where
the \Ds~meson is reconstructed through the four decay modes
$\Dsm \ra \phipi$, $\Dsm \ra \kstark$, $\Dsm \ra \ksk$, and 
$\Dsm \ra \phil$. 
We obtain
\begin {equation}
  \tau(\Bsm) = (1.36\ \pm 0.09 \ ^{+0.06}_{-0.05})\  {\rm ps},
\end {equation}
where the first error is statistical and the second systematic.
This is currently the world's best measurement of the \Bs\ lifetime from
a single experiment. This result agrees with an earlier
CDF measurement~\cite{old_bs_cdf}, which is superseded by the present
measurement.
We determine the $\Bsm/B^0$ lifetime ratio to be $0.899\pm0.072$ 
using the CDF average $B^0$ lifetime~\cite{fumiprd}.

In addition, we have examined
the \Bs\ meson pseudo-proper decay length distribution 
for a lifetime difference $\dgog$ between the two mass eigenstates of the
\Bs\ meson, $\Bsh$ and~$\Bsl$.
Using all four \Ds\ decay modes, an upper limit of $\dgog < 0.83$ is
set at 95\% CL, corresponding to the Standard Model limit
\begin {equation}
  \dms < 96~{\rm ps}^{-1} 
\times \left(\frac{5.6\cdot 10^{-3}}{\dgam/\dm}\right)
\times \left(\frac{1.55~{\rm ps}}{\tau_{\rm m}(\Bsm)}\right)
 \ \ \ (95\%\ {\rm CL}).
\end {equation}
With considerably increased statistics in the next run of the
Tevatron Collider, our sensitivity on
the lifetime difference $\dgog$ will be significantly improved~\cite{tdr}.

\subsection*{Acknowledgments}

     We thank the Fermilab staff and the technical staffs of the
participating institutions for their vital contributions.  
It is a pleasure to thank G.~Buchalla, I.~Dunietz and H.-G.~Moser for
valuable discussions.
This work was
supported by the U.S. Department of Energy and National Science Foundation;
the Italian Istituto Nazionale di Fisica Nucleare; the Ministry of Education,
Science and Culture of Japan; the Natural Sciences and Engineering Research
Council of Canada; the National Science Council of the Republic of China;
the Swiss National Science Foundation; and the A. P. Sloan Foundation.



\begin{table}
\begin{center}
\begin{tabular}{cccccc}
 \Ds\ Decay Mode & $N(\Dsm)$ & $f_{\dsd}$ & $f_{\dsds}$ & 
 $N_{\rm evt}$ & $c\tau(\Bsm)$ \\
\hline
 $\phipi$  &  $220 \pm 21$  &  2.6\,\%  &  0.8\,\% & 350  
	& $418\ _{-39}^{+43}\ \mu$m  \\
 $\kstark$  & $125 \pm 20$  &  2.5\,\%  &  0.8\,\% & 820 
	& $411\ _{-66}^{+73}\ \mu$m \\
 $\ksk$  & $33 \pm 8$  &  1.8\,\%  &  0.6\,\%  & 146 
	&  $397\ _{-152}^{+161}\ \mu$m  \\
 $\phil$  &  $205 \pm 38$  &  5.7\,\%  &  1.7\,\%  & 635
	& $399\ _{-45}^{+50}\ \mu$m  \\
\end{tabular}
\caption[]{Summary of results for the four \Ds\ decays: \\
	-- the number $N(\Dsm)$ of \Ds\ signal events as input to the
	   lifetime fit; \\
	-- the expected fraction $f_{\dsd}$ of 
	   $B\ra D_{\mbox{\sf s}}^{(*)}D^{(*)}$ decays; \\
	-- the expected fraction $f_{\dsds}$ of 
 	   $\Bsm\ra D_{\mbox{\sf s}}^{(*)}D_{\mbox{\sf s}}^{(*)}$ decays; \\
	-- the number $N_{\rm evt}$ of events in the signal samples; \\
	-- the fitted \Bs\ lifetimes $c\tau(\Bsm)$, where the errors
	shown are statistical only.} 
\label{numds}
\end{center}
\end{table}

\begin{table}[tbp]
\begin{center}
\begin{tabular}{lc}
 Error Source & $\Delta c\tau(\Bsm)$  \\
\hline
 Background treatment & $\pm 11$ $\mu$m \\ 
 Non-\Bs\ backgrounds & $\pm 6$ $\mu$m \\ 
 Decay length requirement & $^{+1}_{-5}$ $\mu$m \\ 
 Momentum estimate & \\ 
 \ \ \ \ \ Lepton $\Pt$ dependence & $^{+6}_{-3}$ $\mu$m \\ 
 \ \ \ \ \ $B$ decay model & $^{+3}_{-1}$ $\mu$m \\ 
 \ \ \ \ \ $b$ quark $\Pt$ spectrum & $\pm 5$ $\mu$m \\ 
 \ \ \ \ \ Electron selection & $\pm 3$ $\mu$m \\ 
 Decay length resolution & $^{+7}_{-2}$ $\mu$m \\
 $D^-$ reflection & $\pm 1$ $\mu$m \\ 
 Detector alignment & $\pm 2$ $\mu$m \\ 
\hline
 Total & $^{+17}_{-15}$ $\mu$m \\ 
\end{tabular}
\caption{Compilation of systematic uncertainties in the measurement of the
\Bs\ lifetime 
combining all four \Ds\ decay modes.} 
\label{systematics}
\end{center}
\end{table}

\end{document}